\documentclass[a4paper,fleqn,usenatbib]{mnras}

\usepackage[T1]{fontenc}
\usepackage{ae,aecompl}
\usepackage{xcolor}

\usepackage{graphicx}	
\usepackage{amsmath}	
\usepackage{amssymb}	
\usepackage{pdflscape}	
\usepackage[english]{babel}

\pdfoutput=1

\newcommand{\degrees}{^{\circ}}
\newcommand{\msol}{M_{\rm \odot}}
\newcommand{\lsol}{L_{\rm \odot}}

\newcommand{\alphacen}{$\alpha$ Cen }

\title[Milankovitch Cycles in Binary Systems]{Milankovitch Cycles of Terrestrial Planets in Binary Star Systems}
\author[Duncan Forgan]{Duncan Forgan $^{1}$\thanks{E-mail:dhf3@st-andrews.ac.uk}  \\
$^{1}$Scottish Universities Physics Alliance (SUPA), School of Physics and Astronomy, University of St Andrews, North Haugh, KY16 9SS \\
}

\begin{document}

\date{Accepted}

\pagerange{\pageref{firstpage}--\pageref{lastpage}} \pubyear{}

\maketitle

\label{firstpage}

\begin{abstract}

\noindent The habitability of planets in binary star systems depends not only on the radiation environment created by the two stars, but also on the perturbations to planetary orbits and rotation produced by the gravitational field of the binary and neighbouring planets.  Habitable planets in binaries may therefore experience significant perturbations in orbit and spin.  The direct effects of orbital resonances and secular evolution on the climate of binary planets remain largely unconsidered.

We present latitudinal energy balance modelling of exoplanet climates with direct coupling to an N Body integrator and an obliquity evolution model.  This allows us to simultaneously investigate the thermal and dynamical evolution of planets orbiting binary stars, and discover gravito-climatic oscillations on dynamical and secular timescales.

We investigate the Kepler-47 and Alpha Centauri systems as archetypes of P and S type binary systems respectively.  In the first case, Earthlike planets would experience rapid Milankovitch cycles (of order 1000 years) in eccentricity, obliquity and precession, inducing temperature oscillations of similar periods (modulated by other planets in the system).  These secular temperature variations have amplitudes similar to those induced on the much shorter timescale of the binary period.

In the Alpha Centauri system, the influence of the secondary produces eccentricity variations on 15,000 year timescales.  This produces climate oscillations of similar strength to the variation on the orbital timescale of the binary.  Phase drifts between eccentricity and obliquity oscillations creates further cycles that are of order 100,000 years in duration, which are further modulated by neighbouring planets.

\end{abstract}


\section{Introduction}

Approximately half of all solar type stars reside in binary systems \citep{Duquennoy1991,Raghavan2010}.  Recent exoplanet detections have shown that planet formation in these systems is possible.  Planets can orbit one of the stars in the so-called S type configuration, such as $\gamma$ Cephei \citep{Hatzes2003} HD41004b \citep{Zucker2004} and GJ86b \citep{Queloz2000}.  If the binary semimajor axis is sufficiently small, then the planet can orbit the system centre of mass in the circumbinary or P type configuration.  Planets in this configuration were first detected around post-main sequence stars, in particular the binary pulsar B160-26 \citep{Thorsett1993,Sigurdsson2003}.  The Kepler space telescope has been pivotal in detecting circumbinary planets orbiting main sequence stars, such as Kepler-16 \citep{Doyle2011}, Kepler-34 and Kepler-35 \citep{Welsh2012}, and Kepler-47 \citep{Orosz2012}.

Planets in binary systems are sufficiently common that we should consider their habitability seriously.  As of July 2016, 112 exoplanets have been detected in binary star systems\footnote{http://www.univie.ac.at/adg/schwarz/multiple.html}, giving an occurrence rate of around 4\% (previous estimates on a much smaller exoplanet population by \citealt{Desidera2007} placed the fraction of planets in S type systems at 20\%).  At gas giant masses, the occurrence rate of planets around P type binaries is thought to be similar to that of single stars \citep{Armstrong2014}.

However, theoretical modelling indicates that the dynamical landscape of the binary significantly affects the planet formation process, both for S-type \citep{Wiegert1997,Quintana2002,Quintana2007,Thebault2008,Thebault2009,Xie2010,Rafikov2014,Rafikov2014a} and P-type systems \citep{Doolin2011,Rafikov2013,Martin2013,Marzari2013,Dunhill2013, Meschiari2014,Silsbee2015}.  Therefore, when considering the prospects for habitable worlds in the Milky Way, one must take care to consider the effects that companion stars will have on the thermal and gravitational evolution of planets and moons.

The habitable zone (HZ) concept \citep{Huang1959,Hart_HZ} is often employed to determine whether a detected exoplanet might be expected to be conducive to surface liquid water (that is, if its mass and atmospheric composition allow it).   Initially calculated for the single star case using 1D radiative transfer modelling of the layers of an Earthlike atmosphere \citep{Kasting_et_al_93}, this quickly establishes a range of orbital distances that produce clement planetary conditions.  Over time, line radiative transfer models have been refined, leading to improved estimates of the inner and outer habitable zone edges \citep{Kopparapu2013,Kopparapu2014}.  

In the case of multiple star systems, the presence and motion of extra sources of gravity and radiation have two important effects:

\begin{enumerate}
\item The morphology and location of the system's HZ changes with time, and
\item Regions of the system are orbitally unstable
\end{enumerate}

\noindent These joint thermal-dynamical constraints on habitability have been addressed in a largely decoupled fashion using a variety of analytical and numerical techniques.  

The thermal time dependence of the HZ can be evaluated by combining the flux from both stars, taking care to weight each contribution appropriately, and applying the single star constraints to determine whether a particular spatial location would receive flux conducive to surface water.  \citet{Kane2013} use the aggregate flux to find a peak wavelength of emission.  Assuming the combined spectrum resembles a blackbody, Wien's Law provides an effective temperature for the total insolation, and hence the limits of \citet{Kopparapu2013} can be applied. This approximation is acceptable for P type systems, where the distance from each star to the planet is similar.   

\citet{Haghighipour2013} and \citet{Kaltenegger2013} weight each star's flux by its effective temperature, and then determine the regions at which this weighted flux matches that of a 1 $\msol$ star at the habitable zone boundaries.  This approach is suitable for both S type and P type systems.  A detailed analytic solution for calculations of this nature has been undertaken by \citet{Cuntz2014}.

\citet{Mason2013} take a similar approach, but they also note that for P type systems, the tidal interaction between primary and secondary can induce rotational synchronisation, which can reduce extreme UV flux and stellar wind pressure, improving conditions in the habitable zone compared to the single star case (see also \citealt{Zuluaga2016}).

The dynamical constraints on habitability rely heavily on N Body simulation, most prominently the work of Dvorak \citep{Dvorak1984,Dvorak1986} and \citet{Holman1999a}.  By integrating an ensemble of test particles in a variety of orbits around a binary, regions of dynamical instability can be determined.  \citet{Holman1999a} used these simulations to develop empirical expressions for a critical orbital semimajor axis, $a_c$.  In the case of a P type system, this represents a minimum value - anything inside $a_c$ is orbitally unstable, as given by the following expression:
 
\begin{multline}
a_p > a_c = a_{bin}\left((1.6 \pm 0.04) + (5.1 \pm 0.05) e_{bin} \right. \\
\left.  + (4.12 \pm 0.09) \mu - (2.22 \pm 0.11) e^2_{bin} - (4.27 \pm 0.17) \mu e_{bin}\right. \\
\left.   - (5.09 \pm 0.11) \mu^2 + (4.61 \pm 0.36) \mu^2 e^2_{bin}\right). \label{eq:stable_amin}
\end{multline}

\noindent In the case of an S-type system, $a_c$ represents a maximum value:

\begin{multline}
a_p < a_c = a_{bin}\left( (0.464 \pm 0.006) - (0.38 \pm 0.01) \mu \right. \\
\left. - (0.631 \pm 0.034)  e_{bin}  + (0.586 \pm 0.061) \mu e_{bin} \right. \\
\left.  +(0.15 \pm 0.041) e^2_{bin} - (0.198 \pm 0.074) \mu e^2_{bin} \right) \label{eq:stable_amax}
\end{multline}

\noindent where $a_{bin}$ is the binary semimajor axis, $e_{bin}$ is the binary orbital eccentricity, and $\mu$ represents the binary mass ratio:

\begin{equation}
\mu = \frac{M_2}{M_1+M_2}
\end{equation}  

\noindent The majority of binary habitability calculations rely on the above dynamical constraints.  Notable exceptions include \citet{Eggl2012}'s use of Fast Lyapunov Indicators for chaos detection, which yield slightly smaller values of $a_c$ for S type systems \citep{Pilat-Lohinger2002}, and \citet{Jaime2014}'s use of invariant loops to discover non-intersecting orbits \citep{Pichardo2005}.  There is a good deal of research into spin-orbit alignments of extrasolar planets under the influence of inclined stellar companions (e.g. \citealt{Anderson2016}), but this work rarely pertains to terrestrial planet habitability.  On the other hand, the evolution of planetary rotation period has been studied intently with regards to habitability of planets in single star systems \citep[e.g.][]{Bolmont2014,Brown2014,Cuartas-Restrepo2016}.

All the above approaches to determining habitability in binary systems rely on an initial 1D calculation of the atmosphere's response to radiative flux, where the key dimension is atmospheric depth.  Equally, 1D approaches can consider the latitudinal variation of flux on a planet's surface, giving rise to the so-called latitudinal energy balance models or LEBMs, which have been used both in the single star case \citep{Spiegel_et_al_08,Dressing2010, Vladilo2013} and for multiple stars \citep{Forgan2012,Forgan2014}.  These are better suited to capture processes that depend on atmospheric circulation, such as the snowball effect arising from ice-albedo feedback \citep{Pierrehumbert2005, Tajika2008}, which is likely to occur in systems where the orbits undergo Milankovitch cycles and other secular evolution \citep{Spiegel2010}.  


However, all these approaches typically decouple the thermal from the dynamical.  The orbital constraints on the HZ are considered separately from the radiative transfer calculations.  While they are eventually combined, the binary habitable zones that are constructed do not incorporate the effects of coupled gravito-thermal perturbations.  Indeed, \citet{Holman1999a} admit that their empirical limits on semimajor axis ignore the potential for stable resonances inside the instability region, as well as unstable resonances in stable regions (cf \citealt{Chavez2014}).  It is likely that planets on stable orbits in binary systems will experience relatively strong orbital element evolution.  For example, circumbinary planets can undergo rapid precession of periapsis, which affects their ability to be detected via transit \citep{Kostov2014,Welsh2014}.  Presumably the spin evolution of planets in this situation can proceed with similar rapidity.  Crucially, climate systems are nonlinear, and can alter their state on very short timescales compared to the planet's orbital period.

In this work, we consider coupled gravito-thermal perturbations on the climate of exoplanets in binary systems.  To do so, we present a LEBM directly coupled to an N-Body integrator and an obliquity evolution model.  We use this combined code to investigate the spin-orbital-climate dynamics of putative planets in two archetypal binary systems: the P-type system Kepler-47, a multi-planet circumbinary system which possesses one exoplanet inside the habitable zone \citep{Orosz2012}; and Alpha Centauri, the nearest star system to the Sun, an S type binary system which \emph{was} thought to possess a short period, Earth-mass exoplanet \citep{Dumusque2012}\footnote{This detection is no longer considered to be credible by some groups, due to concerns with how stellar activity is filtered out of radial velocity data \citep{Hatzes2013}.  Recent attempts to detect \alphacen Bb via transit show a null result \citep{Demory2015}, and re-analysis of the radial velocity data suggests that \alphacen Bb does not exist \citep{Rajpaul2015}.}.  By evolving the orbits of the bodies in tandem with the climate, we are able to detect climate variations that are directly linked to the binary's orbit, and the secular evolution of the planet's orbit and spin.

In section \ref{sec:Method}, we describe the LEBM, and how the N Body model is coupled to it.  In section \ref{sec:Results} we describe the simulation setup and results on dynamical and secular timescales, in section \ref{sec:Discussion} we discuss the implications for habitability, and in section \ref{sec:Conclusions} we summarise the work.

\section{Method} \label{sec:Method}
\subsection{Latitudinal Energy Balance Modelling} \label{sec:LEBM}

\noindent Typically, LEBMs solve the following diffusion equation:

\begin{equation} 
C \frac{\partial T(x,t)}{\partial t} - \frac{\partial }{\partial x}\left(D(1 - x^2)\frac{\partial T(x,t)}{\partial x}\right) = S(1-A(T)) - I(T). \label{eq:LEBM}
\end{equation}

\noindent Where $T(x,t)$ is the surface temperature, $C$ is the effective heat capacity of the atmosphere, $S$ is the  insolation flux, $I$ is the IR cooling and $A$ is the albedo.  In the above equation, $C$, $S$, $I$ and $A$ are functions of $x$ (either explicitly, as $S$ is, or implicitly through $T$).   The latitude $\lambda$ appears through $x\equiv \sin \lambda$.   This equation is evolved with the boundary condition $\frac{dT}{dx}=0$ at the poles (where $\lambda=[-90,90] \degrees$), and 
requires the assumption that the planet rotates rapidly relative to its orbital period.  Our implementation of the LEBM follows that of \citet{Spiegel_et_al_08}, and has been used previously in studying the climate evolution of planets in binary systems on timescales of order a few hundred years \citep{Forgan2012,Forgan2014}.  In our approach, we consider a given latitude to be habitable if its temperature resides within $273 K < T < 373$ K, i.e. that surface water is liquid.  

The diffusion coefficient $D$ determines the efficiency of heat redistribution across latitudes. Its value is defined such that a fiducial Earthlike planet, rotating with period 1 day, orbiting at 1 au around a star of $1 \msol$, produces the correct average temperature profile (see e.g. \citealt{Spiegel_et_al_08, Vladilo2013}).  If the planet's rotation is more rapid, the Coriolis effect will inhibit latitudinal heat transport (see \citealt{Farrell1990}):

\begin{equation} 
D=5.394 \times 10^2 \left(\frac{\Omega_{rot}}{\Omega_{rot,\oplus}}\right)^{-2},\label{eq:D}
\end{equation}

\noindent where $\Omega_{rot}$ is the rotational angular velocity of the planet, and $\Omega_{rot,\oplus}$ is the rotational angular velocity of the Earth.   This is a necessarily simple expression, but can be made more rigorous through including terms for atmospheric pressure and mean molecular weight (e.g. \citealt{Williams1997a}, but see also \citealt{Vladilo2013}'s attempts to introduce a latitudinal dependence to $D$ to mimic the Hadley convective cells on Earth).  Beyond this, full global circulation modelling is needed to explore the effects of rotation \citep{DelGenio1993,DelGenio1996}.  

As in previous work, we solve equation (\ref{eq:LEBM}) using an explicit forward time, centre space finite difference algorithm.  A global timestep is used, with standard constraint

\begin{equation}
\Delta t_{LEBM} < \frac{\left(\Delta x\right)^2C}{2D(1-x^2)}.  
\end{equation}

\noindent The atmospheric heat capacity $C$, is a function of the planet's surface ocean fraction and how much of that is frozen, $f_{ice}$:

\begin{equation} 
C = f_{land}C_{land} + f_{ocean}\left((1-f_{ice})C_{ocean} + f_{ice} C_{ice}\right),
\end{equation}

\noindent where $f_{land}=1-f_{ocean}$.  The heat capacities of land, ocean and ice covered areas are

\begin{align*} 
C_{land} & =5.25 \times 10^9  \mathrm{erg \, cm^{-2}\, K^{-1}} \\
C_{ocean} & = 40.0C_{land} \\
C_{ice} & = \left\{
\begin{array}{l l }
9.2C_{land} & \quad \mbox{263 K $< T <$ 273 K} \\
2C_{land} & \quad \mbox{$T<263$ K}. \\
\end{array} \right. 
\end{align*}

\noindent The infrared cooling function $I$ is 

\begin{equation} 
I(T) = \frac{\sigma_{SB}T^4}{1 +0.75 \tau_{IR}(T)}, 
\end{equation}

\noindent with the optical depth of the atmosphere given as

\begin{equation} 
\tau_{IR}(T) = 0.79\left(\frac{T}{273\,\mathrm{K}}\right)^3. 
\end{equation}

\noindent The albedo function is

\begin{equation} 
A(T) = 0.525 - 0.245 \tanh \left[\frac{T-268\,\mathrm K}{5\, \mathrm K} \right].
\end{equation}

\noindent This correctly reproduces the ice-albedo feedback phenomenon, which allows a rapid non-linear increase in albedo as the ice coverage increases.

At any instant, for a single star, the insolation received at a given latitude at an orbital distance $r$ is

\begin{equation}
S = q_0\cos Z \left(\frac{1 AU}{r}\right)^2,
\end{equation}

\noindent where $q_0$ is the bolometric flux received from the star at a distance of 1 AU, and $Z$ is the zenith angle:

\begin{equation} q_0 = 1.36\times 10^6\left(\frac{M}{\msol}\right)^4
  \mathrm{erg \,s^{-1} cm^{-2}} 
  \end{equation}

\begin{equation} \cos Z = \mu = \sin \lambda \sin \delta + \cos
  \lambda \cos \delta \cos h. \end{equation} 

The solar hour angle is $h$, and $\delta$ is the solar declination, which is calculated by computing the scalar product of the spin-axis vector $\mathbf{s}$ and the planet-star separation vector $\mathbf{r}$.  We obtain the spin-axis vector by rotation of the angular momentum vector in the x-axis by $\delta_0$, followed by a rotation around the axis defined by the angular momentum vector by $p_a$, the axial precession angle (or longitude of winter solstice).  

Our rapid rotation assumption requires that we use diurnally averaged quantities, so we also diurnally average $S$:

\begin{equation} S = q_0 \bar{\mu}. \end{equation}

\noindent We do this by integrating $\mu$ over the sunlit part of the day, i.e. $h=[-H, +H]$, where $H(x)$ is the radian half-day length at a given latitude.  Multiplying by $H/\pi$ (as $H=\pi$ if a latitude is illuminated for a full rotation) gives the total diurnal insolation as

\begin{equation} 
S = q_0 \left(\frac{H}{\pi}\right) \bar{\mu} = \frac{q_0}{\pi} \left(H \sin \lambda \sin \delta + \cos \lambda \cos \delta \sin H\right). 
\end{equation}

\noindent The radian half day length is calculated as

\begin{equation} 
\cos H = -\tan \lambda \tan \delta. 
\end{equation}

\noindent The total insolation is a simple linear combination of the contributions from both stars.  If one star is eclipsed by the other, then we set its contribution to $S$ to zero.  We ensure that the simulation can accurately model an eclipse by adding an extra timestep criterion, ensuring that the transit's duration will not be less than ten timesteps.  

We fix the parameters of the model to those of the Earth: the initial obliquity is set to 23.5 degrees, and the ocean fraction $f_{ocean} = 0.7$.  The rotation period of the body is 1 day.  It is important to note that altering these parameters will alter the strength of climate fluctuations, especially if orbits are eccentric.  Indeed, \citet{Forgan2012} showed that reducing the planet's ocean fraction can significantly boost temperature fluctuations in S-type binary systems with fixed orbits, and that increasing obliquity while holding other parameters fixed typically increases the average temperature of the planet.  The following results should be considered with these facts in mind.

\subsection{The N-Body Model}\label{sec:NBody}

\noindent The dynamical evolution of the system utilises a standard 4th-order Hermite integrator with an adaptive shared timestep.  We calculate this N Body timestep for all bodies$\{i\}$, $\Delta t_{N}$, by finding the minimum value of $\Delta t_{i}$:

\begin{equation} \Delta t_{i} = \left(\eta \frac{\frac{a_i}{j_i}+
    \frac{j_i}{s_i} }{\frac{c_i}{s_i} +
    \frac{s_i}{j_i}}\right)^{1/2} .
\end{equation}

\noindent Here, $a$ represents the magnitude of the body's acceleration, $j_i$ $s_i$ and $c_i$ are the magnitudes of the first, second and third derivatives of the acceleration of particle $i$ respectively, and $\eta$ is a tunable parameter which we set to 0.002.  This is a fairly strict timestep condition, and as such the error in angular momentum is typically one in $10^6$ or better throughout.  

\subsection{Obliquity Evolution}\label{sec:obliq}

\noindent We adopt the obliquity evolution model of \citet{Laskar1986,Laskar1986a}, developed for the Solar System and subsequently used for putative exoplanet systems \citep{Armstrong2004,Armstrong2014a}.  In this paradigm, the evolution of the obliquity $\delta_0$ and precession $p_a$ are functions of the inclination variables

\begin{align}
p &= \sin \left(\frac{i}{2}\right) \sin \Omega \\
q &= \sin \left(\frac{i}{2}\right) \cos \Omega 
\end{align}

\noindent Where $i$ is the inclination, and $\Omega$ is the longitude of the ascending node.  The obliquity and precession evolve according to the following:

\begin{align}
\frac{d\delta_0}{dt} & = -B \sin p_a + A \cos p_a \\
\frac{d p_a}{dt} & = R(\delta_0) - \cot \delta_0 \left( A \sin p_a + B \cos p_a \right) -2C -p_g.
\end{align}

\noindent $A,B$ and $C$ are all functions of $p$ and $q$:

\begin{align}
A(p,q) & = \frac{2}{\sqrt{1-p^2-q^2}}\left(\dot{q} - pC(p,q)\right) \\
B(p,q) & = \frac{2}{\sqrt{1-p^2-q^2}}\left(\dot{p} - qC(p,q)\right) \\
C(p,q) & = \dot{p}q - \dot{q}p
\end{align}

Note that these $A,B,C$ terms ensure increases in inclination mediate changes in obliquity.  Equivalently, if the inclination of a planet's orbit is increased, the obliquity decreases, as the angle between the orbital plane and the fundamental plane defined by the planet's spin axis decreases (see Figure 1 of \citealt{Armstrong2014a}).

That being said, the spin axis of the planet can change regardless of the inclination, due to either direct torques from the star ($R(\delta_0)$) or from the relativistic precession term $p_g$.  \citet{Laskar1986} give the direct torque from a single host star as

\begin{equation}
R(\delta_0) =  \frac{3k^2M_*}{a^3 \Omega_{rot}} E_D S_0 \cos \delta_0
\end{equation}

Where $E_D$ is the dynamical ellipticity (i.e. the non-sphericity) of the planet (which we set equal to 0.00328005 for the remainder of this work), 

\begin{equation}
S_0 = \frac{1}{2}\left(1-e^2\right)^{-3/2} -0.422 \times 10^{-6}
\end{equation}

and $k^2 = \frac{GM_*}{4\pi^2}$ (where the units of $G$ must be selected to be appropriate for comparison with $\Omega_{rot}$).  For a single star, the relativistic precession is

\begin{equation}
p_g = \frac{k_r}{2(1-e^2)}
\end{equation}

\noindent where

\begin{equation}
k_r = \frac{n^3 a^2}{c^2\left(1 + M_p/M\right)}
\end{equation}

\noindent The mean motion $n$ can be determined by considering $k$ in the context of Kepler's third law:

\begin{equation}
n^2 a^3 = \frac{k^2}{\left(1+M_p/M\right)}
\end{equation}

\noindent In this work, we make the following assumptions about these equations in their use for binary stars.  In the S type case, we assume that direct torques and precession is generated by the host star only.  The secondary can influence the obliquity only through modification of the planet's orbital elements $e,i,\Omega$.

In the case of a P type system, we assume that the torques from both stars co-add.  The planet's orbital elements relative to the system centre of mass are employed in both cases for simplicity.  Given the distance of both stars from the centre of mass is small relative to the planet's semi-major axis, this seems a reasonable assumption (although we do note the need for further investigation of this problem, see Discussion).

\subsection{Coupling the Models}

\noindent To couple the LEBM to the N Body integrator and obliquity evolution model, we elect the simplest route, by forcing all systems to evolve according to a shared timestep.  In practice, this means comparing the LEBM and N Body timesteps, i.e.

\begin{equation}
\Delta t = min \left(\Delta t_{N}, \Delta t_{LEBM}\right).
\end{equation}

\noindent Typically the obliquity evolution timestep is much larger than the other two.  This does limit the code's efficacy when evolving systems with either short dynamical timescales, or short thermal timescales.  In the case of a fiducial Earth-Sun model, we are able to evolve the coupled LEBM-N Body system with similar runtime to a LEBM using fixed Keplerian orbits.  We will see that in the S type configuration, the addition of N Body physics makes little appreciable difference to computational speed.  However, in the P type configuration, the short dynamical timescale of the binary increases the runtime significantly.  This could be alleviated by other timestepping approaches, which we address in the Discussion.

We emphasise that correctly resolving the LEBM is crucial - it is a nonlinear system, with positive feedback mechanisms that can operate rapidly compared to the system's spin-orbit dynamical time.  It is this property that requires the models to be fully coupled in order to truly understand the climate of planets in dynamically rich systems over secular timescales.

We have tested the N Body integrator and obliquity evolution model against the results of \citet{Armstrong2014a} (their System 1), and find a good match for their orbital elements and spin parameters.  In a companion paper (Forgan and Mead, in prep) we test the spin-orbit-climate evolution of the Earth under the influence of the Solar System planets, and find that appropriate Milankovitch cycles in the planet's spin-orbit parameters do indeed arise. 

\section{Results} \label{sec:Results}

\noindent We now apply our combined model to the two archetypal P and S type binary systems.  We will be comparing runs with obliquity evolution switched on and off to investigate what climate features are due to either orbital or spin evolution.

\subsection{Kepler-47}

\subsubsection{Setup}

\noindent The Kepler-47 system contains a 1.043 $\msol$ star and an 0.362 $\msol$ star orbiting each other with a period of around 7.5 days.  We adopt the orbital parameters of \citet{Orosz2012}, with a semi-major axis of 0.0836 AU and eccentricity 0.0234, and assume that the stars' luminosities are determined by standard main sequence relations.   

Kepler-47c orbits inside the circumbinary habitable zone at 0.989 AU, with an eccentricity upper limit of 0.41.  As we are using the Kepler-47 system as an archetype for terrestrial habitability in P type systems, we replace Kepler-47c with an Earth mass planet orbiting at the same semi-major axis, and investigate both low and high eccentricity orbits.  Kepler-47b orbits interior to Kepler-47c with a semimajor axis of 0.2956 AU with eccentricity 0.034, and period 49.5 days.  We investigate the climate of our terrestrial planet both with and without Kepler-47b's presence.

\subsubsection{Zero Eccentricity, Without Kepler-47b}

\begin{figure*}
\begin{center}$\begin{array}{cc}
\includegraphics[scale=0.4]{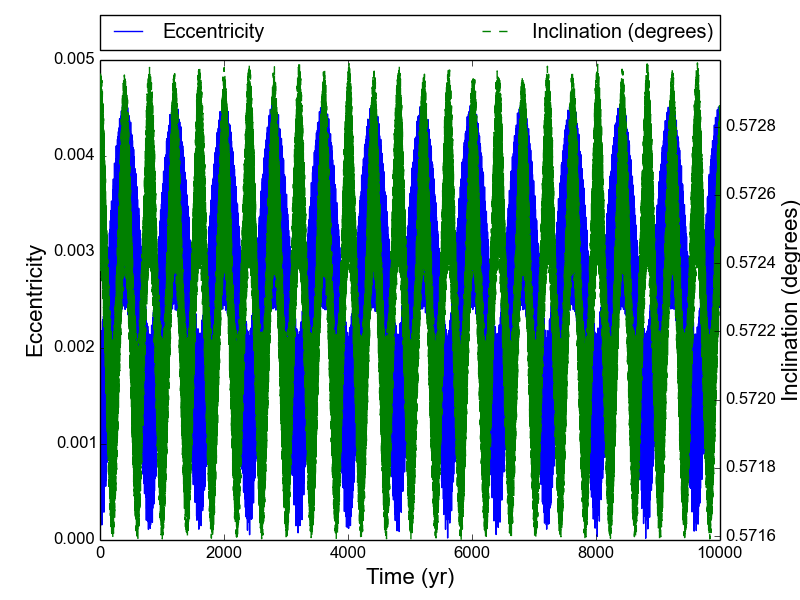} &
\includegraphics[scale=0.4]{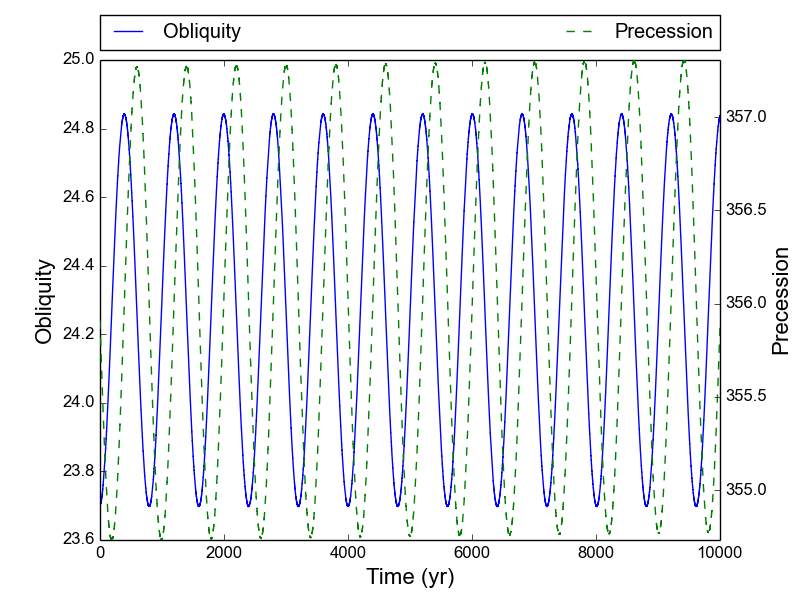}\\
\end{array}$
\caption{The dynamical evolution of the terrestrial planet with Kepler-47c's semimajor axis, and zero eccentricity. Left, the orbital evolution of the body, as given by its eccentricity and inclination.  Right, the spin evolution as given by the obliquity and precession angles. \label{fig:Kepler47c_e0_orbit}}
\end{center}
\end{figure*}

\begin{figure*}
\begin{center}$\begin{array}{cc}
\includegraphics[scale=0.4]{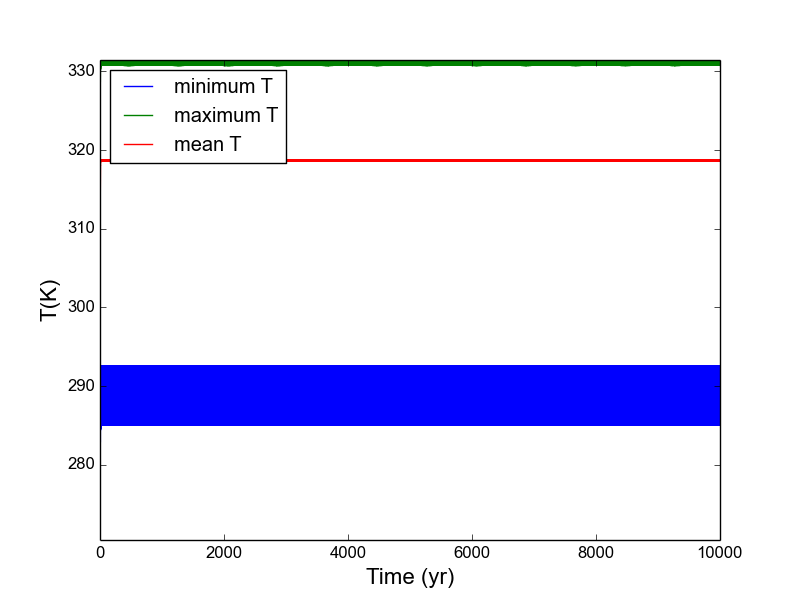} &
\includegraphics[scale=0.4]{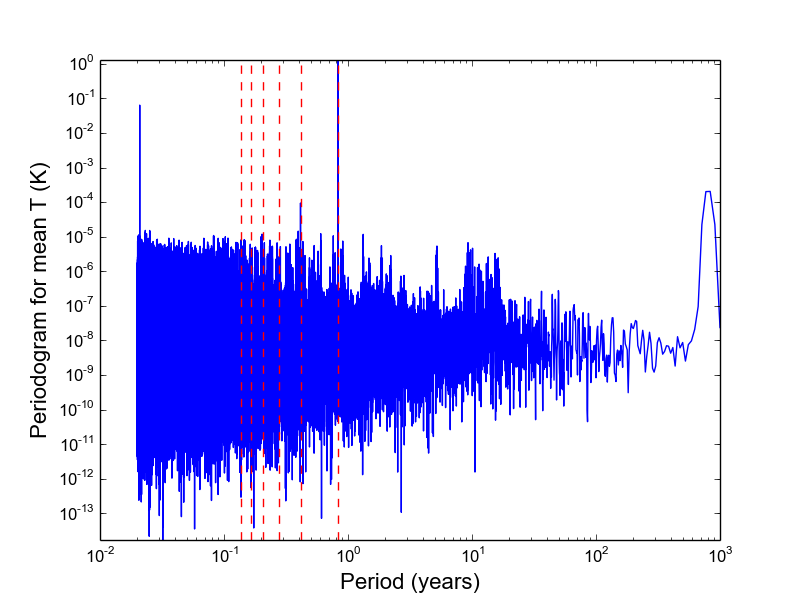} \\
\includegraphics[scale=0.4]{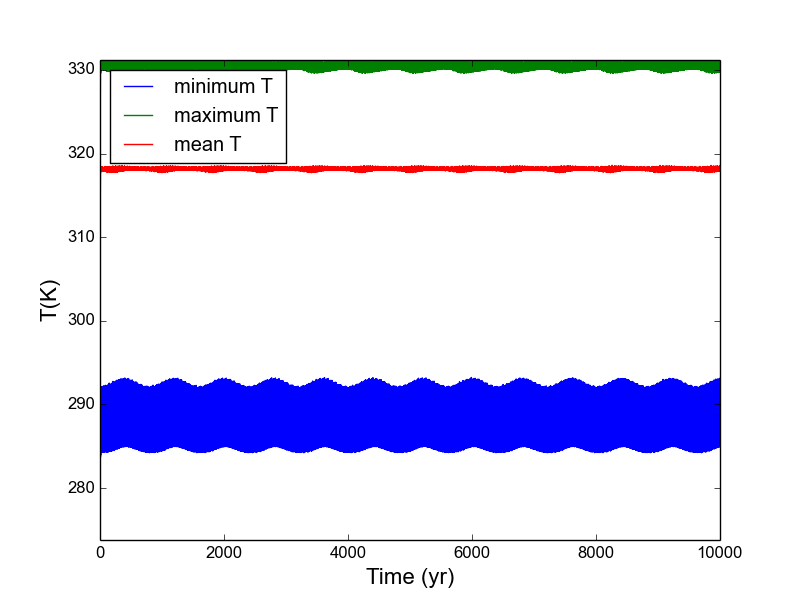} &
\includegraphics[scale=0.4]{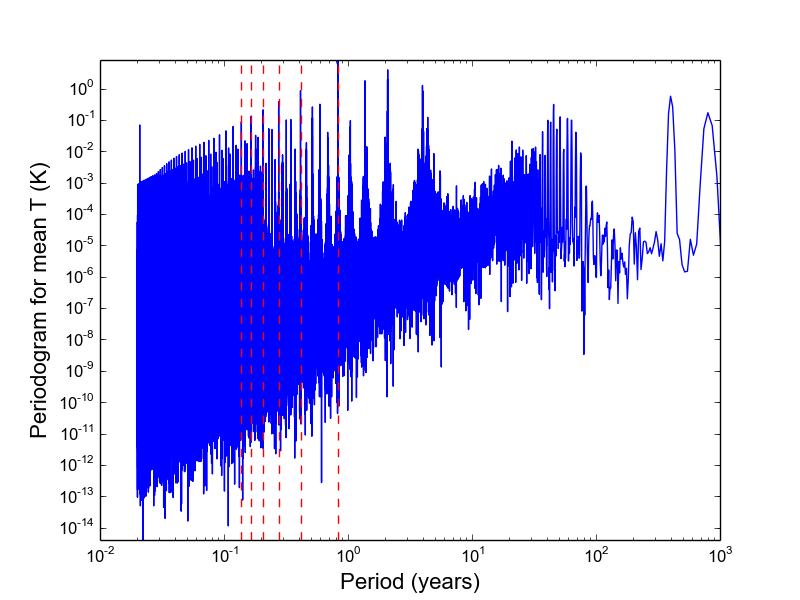} \\
\end{array}$
\caption{The climate evolution of the Kepler-47c terrestrial planet, with obliquity evolution switched off (top row) and switched on (bottom row).  Left: The global maximum, minimum and mean temperatures on the surface over 10,000 years.  Right:  Periodograms for the mean temperature.  The red dashed lines indicate the planet's orbital period of 0.829 years, and its harmonics ($1/2$, $1/3$... of the period).  \label{fig:Kepler47c_e0_climate}}
\end{center}
\end{figure*}

\noindent Figure \ref{fig:Kepler47c_e0_orbit} shows the orbital evolution of a terrestrial planet orbiting the Kepler-47 binary at $a_p=0.989$ AU with zero eccentricity and an initial inclination of 0.5$^\circ$ relative to the binary plane.  We run the simulation for 10,000 years, with sufficiently high snapshot frequency that the orbital period of the binary (0.0205 years) is well resolved.  The planet's orbit is relatively stable, undergoing small eccentricity and inclination variations of around 800 and 400 year periods respectively (note also that the argument of periapsis precesses on a similar timescale). 

In the case where the obliquity is fixed, the planet's climate settles to a stable state, with mean temperatures fluctuating by around 0.1 K (top row of Figure \ref{fig:Kepler47c_e0_climate}).  We can see in the periodogram for fixed obliquity that the major contribution to temperature fluctuation is seasonal variation over the orbital period of 0.829 years (and its harmonics at $1/n$ of the period), closely followed by a contribution at the binary period of 0.0205 years as the relative insolation from each object varies.  Finally, we see a significantly weaker contribution from eccentricity variation at 800 years.  There are no low order mean motion resonances between the binary and planet period - the system is closest to a 80:2 resonance.  There is no evidence of such a resonance in the temperature data, which would result in a peak at approximately 1.66 years in the periodogram.

In the case where obliquity is allowed to vary (bottom row of Figure \ref{fig:Kepler47c_e0_climate}), we can immediately detect climatic variations from inspecting the maximum, mean and minimum temperature curves.  The presence of an extra peak at around 400 years in the temperature periodogram (bottom right of Figure \ref{fig:Kepler47c_e0_climate}) shows that the inclination is forcing similar variations in obliquity and precession angle (Figure \ref{fig:Kepler47c_e0_orbit}).  Generally speaking, the planet's climate now shows a richer set of resonant features in the periodogram with periods greater than that of the orbital periods in play.

\subsubsection{Zero Eccentricity, with Kepler-47b}

\noindent The previous section has shown that single planets in P type systems will undergo secular evolution quite similar to that of Milankovitch cycles (albeit at a much reduced timescale).  We now add Kepler-47b to the system (with zero eccentricity and inclination) to gauge what effect neighbouring planets might have on the secular evolution of circumbinary habitable climates.

\begin{figure*}
\begin{center}$\begin{array}{cc}
\includegraphics[scale=0.4]{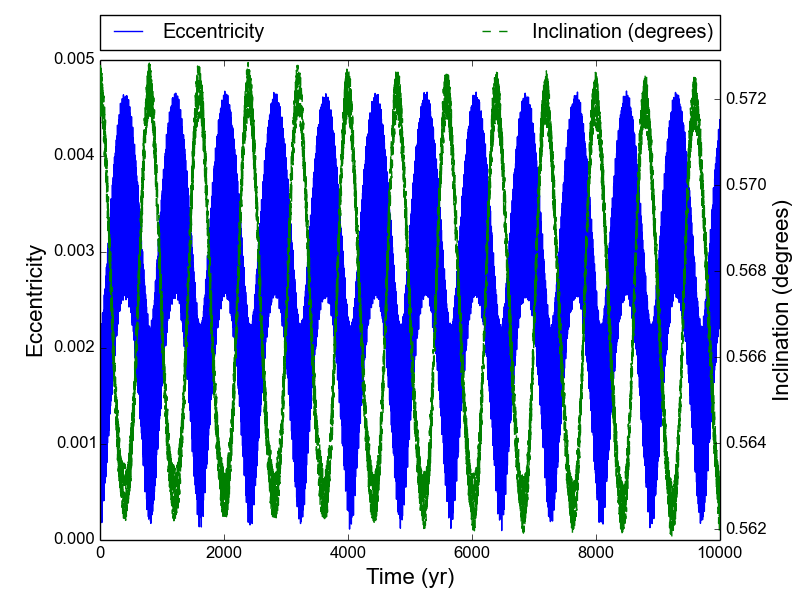} &
\includegraphics[scale=0.4]{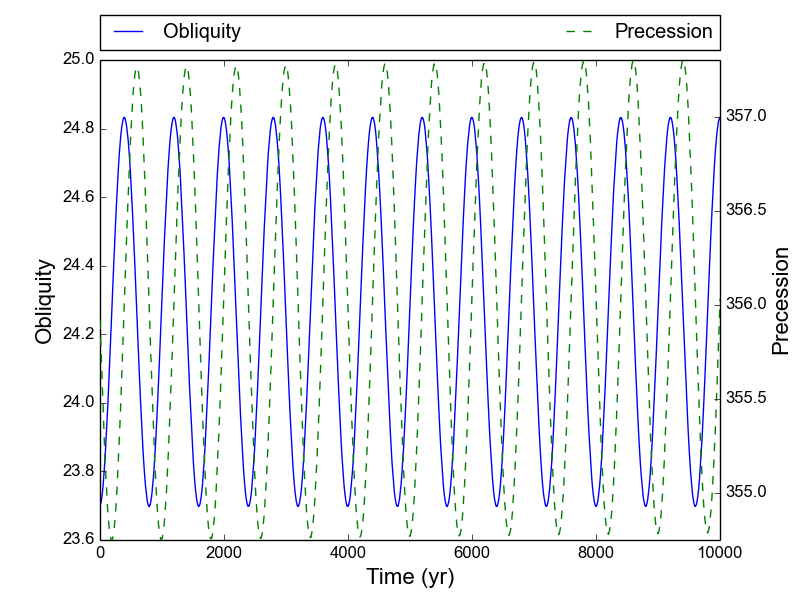}\\
\end{array}$
\caption{The dynamical evolution of the terrestrial planet with Kepler-47c's semimajor axis, and zero eccentricity, in the presence of Kepler-47b. Left, the orbital evolution of the body, as given by its eccentricity and inclination.  Right, the spin evolution as given by the obliquity and precession angles. \label{fig:Kepler47c_b_e0_orbit}}
\end{center}
\end{figure*}

\begin{figure*}
\begin{center}$\begin{array}{cc}
\includegraphics[scale=0.4]{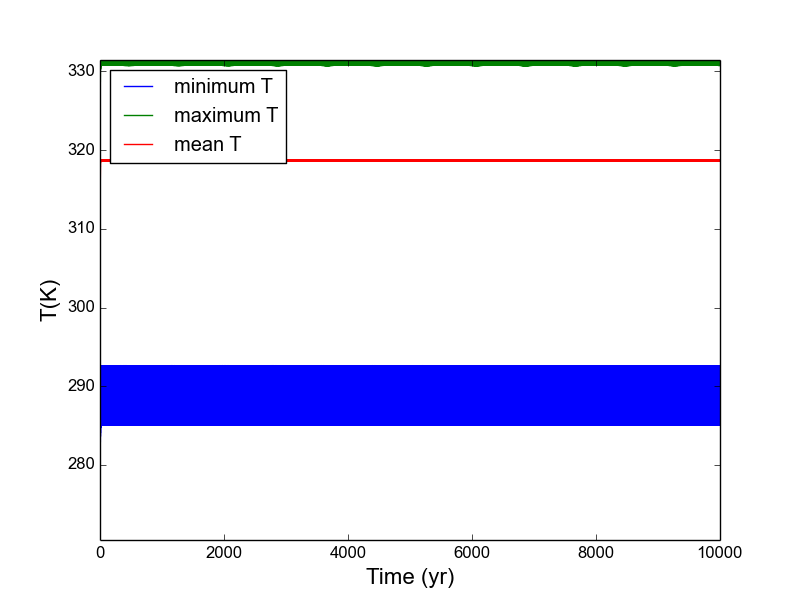} &
\includegraphics[scale=0.4]{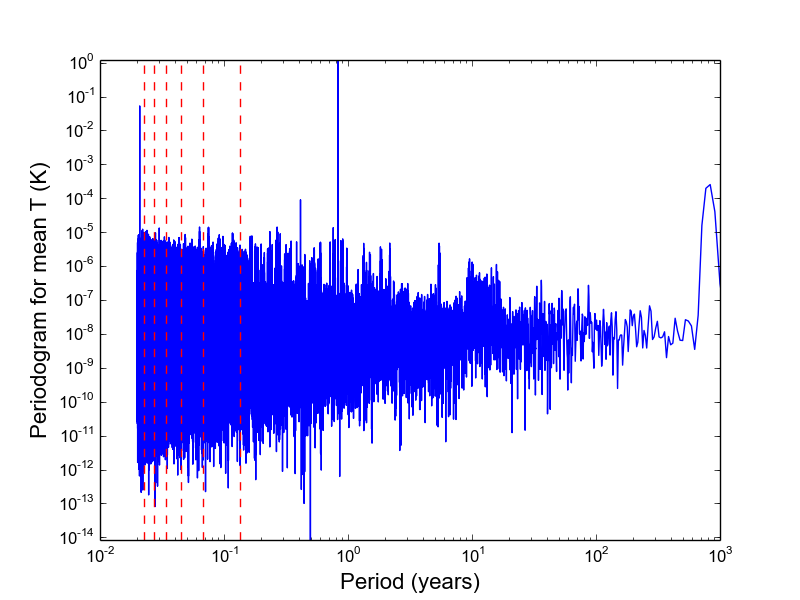} \\
\includegraphics[scale=0.4]{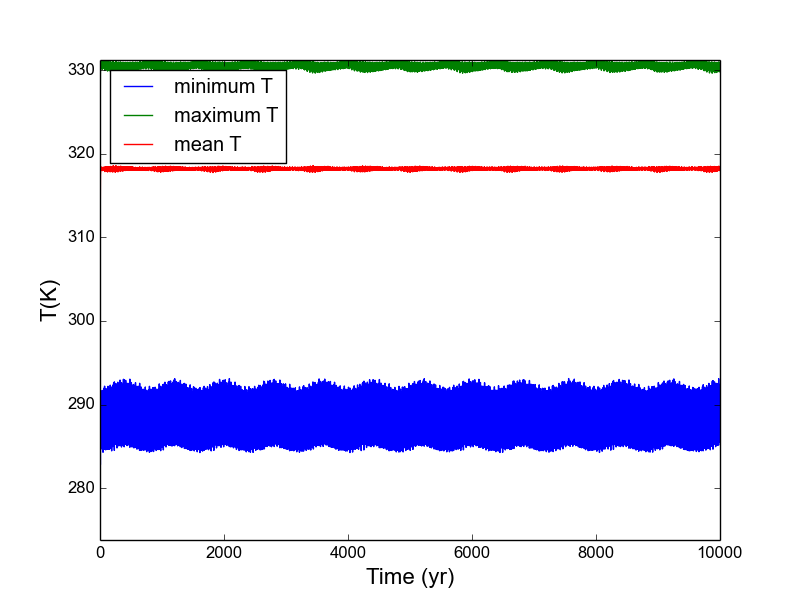} &
\includegraphics[scale=0.4]{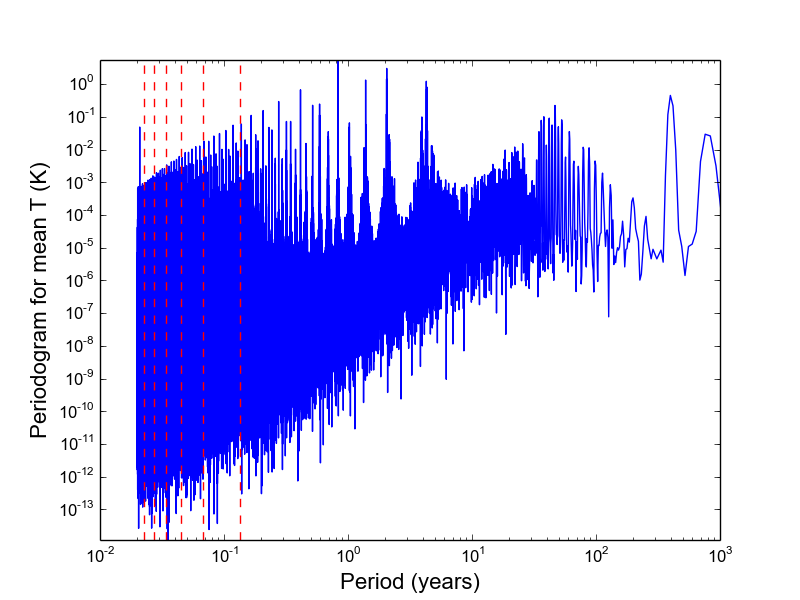} \\
\end{array}$
\caption{The climate evolution of the Kepler-47c terrestrial planet in the presence of Kepler-47b, with obliquity evolution switched off (top row) and switched on (bottom row).  Left: The global maximum, minimum and mean temperatures on the surface over 10,000 years.  Right:  Periodograms for the mean temperature.  The red dashed lines indicate Kepler-47b's orbital period of 0.1355 years, and its harmonics ($1/2$, $1/3$... of the period).  \label{fig:Kepler47c_b_e0_climate}}
\end{center}
\end{figure*}

\noindent Figure \ref{fig:Kepler47c_b_e0_orbit} shows the orbital evolution of the Kepler-47c substitute.  Comparing to the previous section (Figure \ref{fig:Kepler47c_e0_orbit}), we see that the eccentricity variation has not changed much, but the inclination variation has decreased its period by a factor of roughly two.  Interestingly, no such changes are seen in the obliquity and precession evolution, indicating that stellar torques are presumably dominant.

The periodograms for both cases (Figure \ref{fig:Kepler47c_b_e0_climate}) show little change in the climate by adding a neighbour planet.  The periodograms show no signs of Kepler-47b's influence at its orbital period of 0.1355 years.  The features seen at 0.1355 years with obliquity evolution exist in the previous run without Kepler-47b.  The planets are not in mean motion resonance - they are closest to a 49:8 mean motion resonance, which would indicate a peak at approximately 6.63 years, which is not seen in either case.
 

\subsubsection{High Eccentricity, no b}

\begin{figure*}
\begin{center}$\begin{array}{cc}
\includegraphics[scale=0.4]{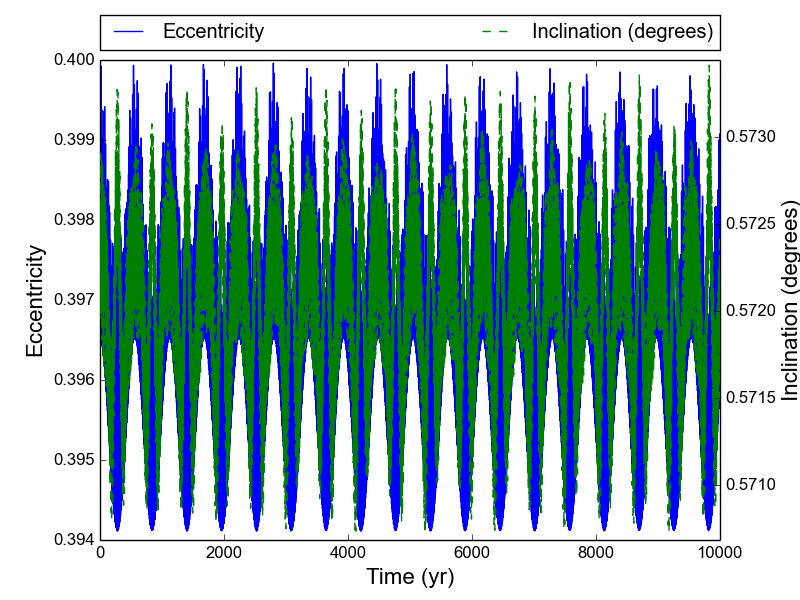} &
\includegraphics[scale=0.4]{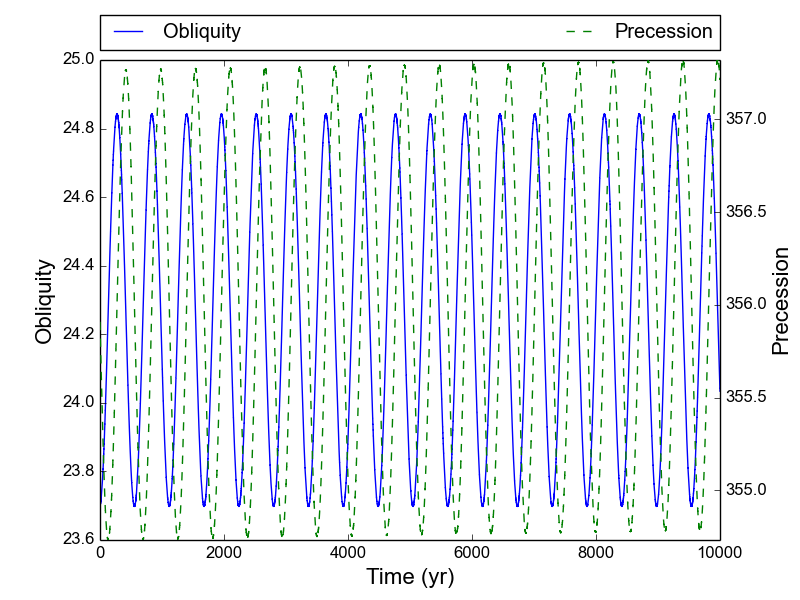}\\
\end{array}$
\caption{The dynamical evolution of the terrestrial planet with Kepler-47c's semimajor axis, and eccentricity 0.4. Left, the orbital evolution of the body, as given by its eccentricity and inclination.  Right, the spin evolution as given by the obliquity and precession angles. \label{fig:Kepler47c_e04_orbit}}
\end{center}
\end{figure*}

We now remove Kepler-47b from the system, and increase the eccentricity of our habitable planet to 0.4.  The dynamical evolution (Figure \ref{fig:Kepler47c_e04_orbit}) is more rapid, with small eccentricity and inclination oscillations about the original value with a period of around 550 years, and similar obliquity and precession evolution.  Note the amplitude modulation of the inclination, which coincides with peak eccentricity.

\begin{figure*}
\begin{center}$\begin{array}{cc}
\includegraphics[scale=0.4]{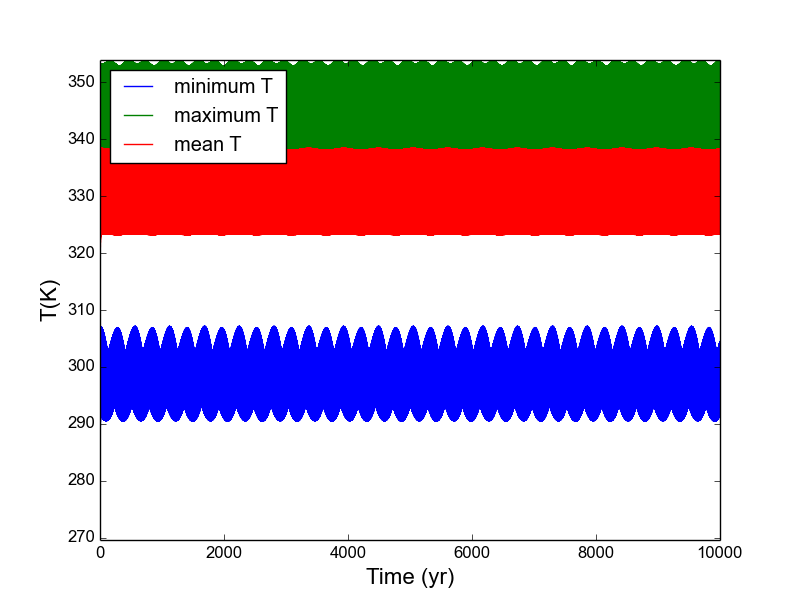} &
\includegraphics[scale=0.4]{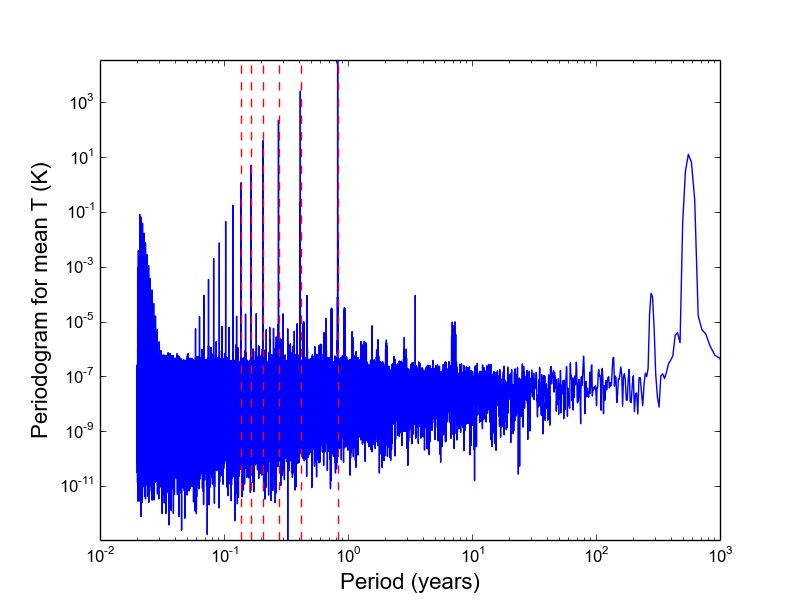} \\
\includegraphics[scale=0.4]{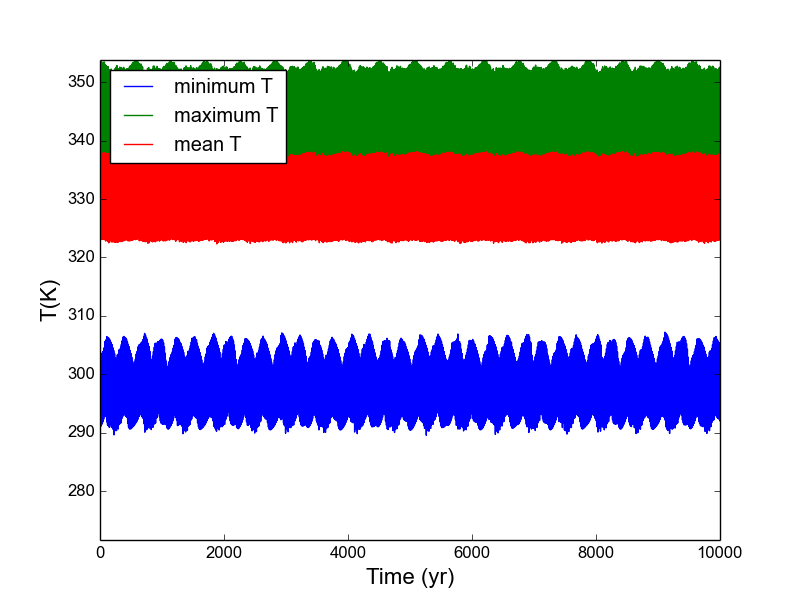} &
\includegraphics[scale=0.4]{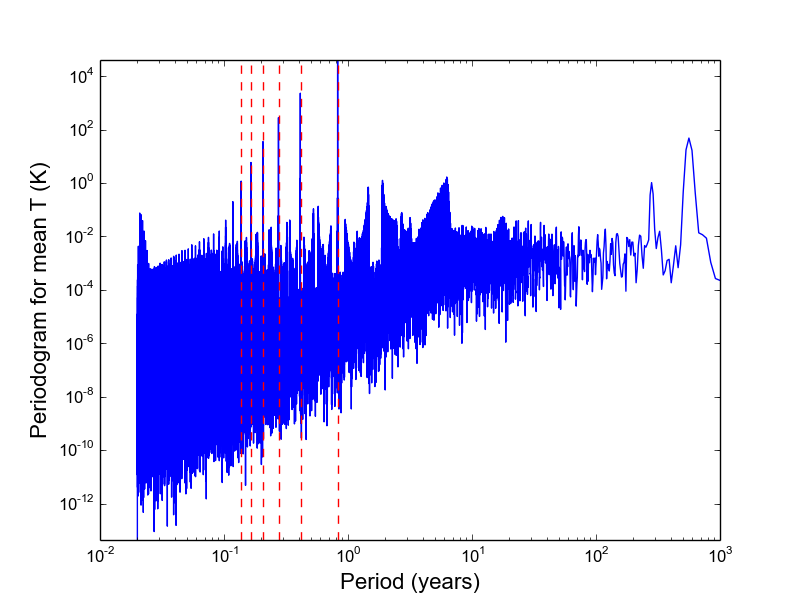} \\
\end{array}$
\caption{The climate evolution of the Kepler-47c terrestrial planet at high eccentricity, with obliquity evolution switched off (top row) and switched on (bottom row).  Left: The global maximum, minimum and mean temperatures on the surface over 10,000 years.  Right:  Periodograms for the mean temperature.  The red dashed lines indicate the planet's orbital period of 0.829 years, and its harmonics ($1/2$, $1/3$... of the period).  \label{fig:Kepler47c_e04_climate}}
\end{center}
\end{figure*}

Naturally, the climate of the body experiences stronger temperature oscillations even with obliquity switched off (top row of Figure \ref{fig:Kepler47c_e04_climate}).  The periodogram shows greater importance for the seasonal variation, as well as the eccentricity variation peak at 550 years.  As the planet and binary are not in mean motion resonance, the contribution of the binary to the planet's eccentricity periodogram is smeared between 0.02 and 0.03 years due to the planet's increased eccentricity.  Note that this increased eccentricity raises the maximum temperature beyond the runaway greenhouse limit of 340K.  The runaway greenhouse effect is not modelled by the LEBM, and we should be careful when making statements about this configuration's habitability.  Some weak modes appear around the planet's orbital period of 0.829 years, but their origin is unclear - presumably they are linked to the precession of the planet's periapsis relative to that of the binary.

Allowing obliquity to vary allows other oscillations to assume greater importance.  Indeed, the variations caused by binary motion are close to negligible in this case, especially compared to variations in the year-decade range.

\subsection{Alpha Centauri B}

\subsubsection{Setup}

\noindent The Alpha Centauri system is in fact a hierarchical triple system, with Alpha Centauri A and B orbiting each other at  23.4 AU with eccentricity 0.5179.  We neglect the third component, Proxima Centauri, as it orbits at great distance and is of sufficiently low mass \citep{Wertheimer2006}.  We consider \alphacen B as the host star for a planetary system.

The stellar masses are $M_A=1.1 \msol$, $M_B = 0.934 \msol$, and their luminosities are $L_A=1.519 \lsol$ and $L_B=0.5 \lsol$ respectively \citep{Thevenin2002}.  This modifies the location of the habitable zone as was previously measured by \citet{Forgan2012}, as they used main sequence relations for the luminosity.

We do not model the presence of \alphacen Bb, as its 3 day orbit would place it extremely close to \alphacen B, and hence is unlikely to produce a significant perturbation on any planets within the habitable zone. Instead, we place a single Earthlike planet in the system near the outer edge of the habitable zone, on a circular orbit at $0.7095$ AU , where the effects of \alphacen A are maximal.  To ensure obliquity evolution occurs, we give our planet a small inclination of $0.5^\circ$ relative to the binary plane.

However, we do wish to consider the relative strength of Milankovitch cycles resulting from the binary compared to those induced by neighbouring planets (cf Figure 8 of \citealt{Andrade-Ines2014}).  We attempt to maximise this effect by running another set of models with a second Earth-mass body orbiting in 3:2 resonance with our habitable world (with a zero inclination orbit). 

\subsubsection{Single Planet Runs}

%

\begin{figure*}
\begin{center}$\begin{array}{cc}
\includegraphics[scale=0.4]{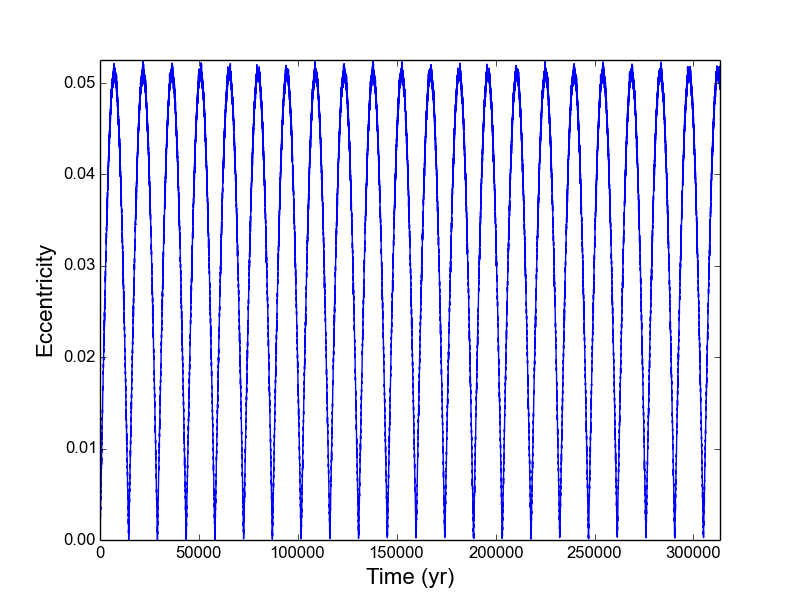} &
\includegraphics[scale=0.4]{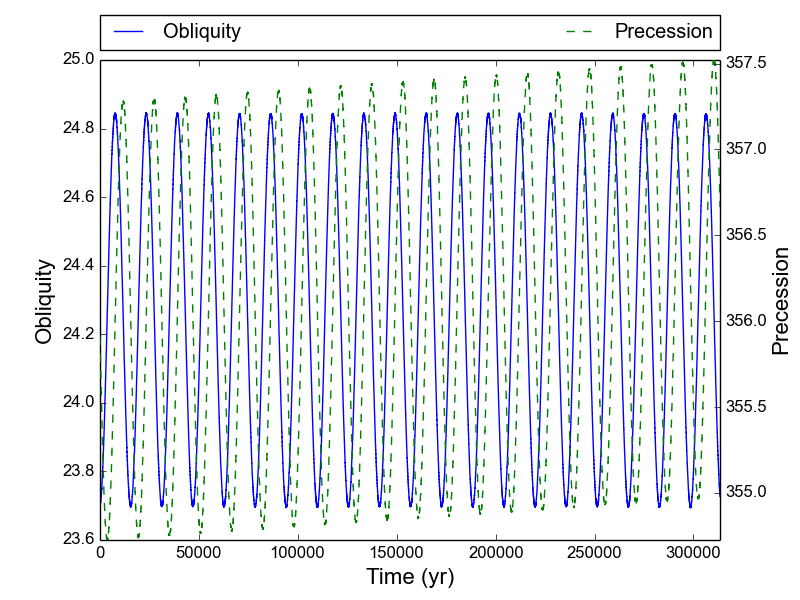}\\
\end{array}$
\caption{The dynamical evolution of the terrestrial planet orbiting \alphacen B. Left, the orbital evolution of the body, as given by its eccentricity.  We refrain from plotting the inclination, as its fluctuations are extremely low with no obvious periodic oscillation.  Right, the spin evolution as given by the obliquity and precession angles. \label{fig:alphacen_e0_orbit}}
\end{center}
\end{figure*}

Figure \ref{fig:alphacen_e0_orbit} shows the dynamical evolution of the planet around \alphacen B.  The initially zero eccentricity is forced to a maximum of 0.05 on a cycle of approximately 14,500 years.  The obliquity and precession evolve with a slightly longer period, resulting in the eccentricity and obliquity cycles drifting in and out of phase.

\begin{figure*}
\begin{center}$\begin{array}{cc}
\includegraphics[scale = 0.4]{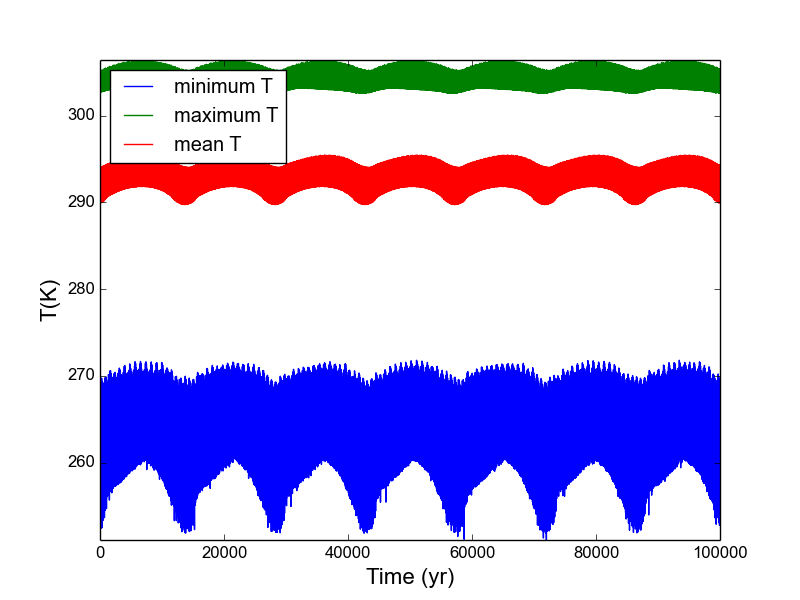} &
\includegraphics[scale=0.4]{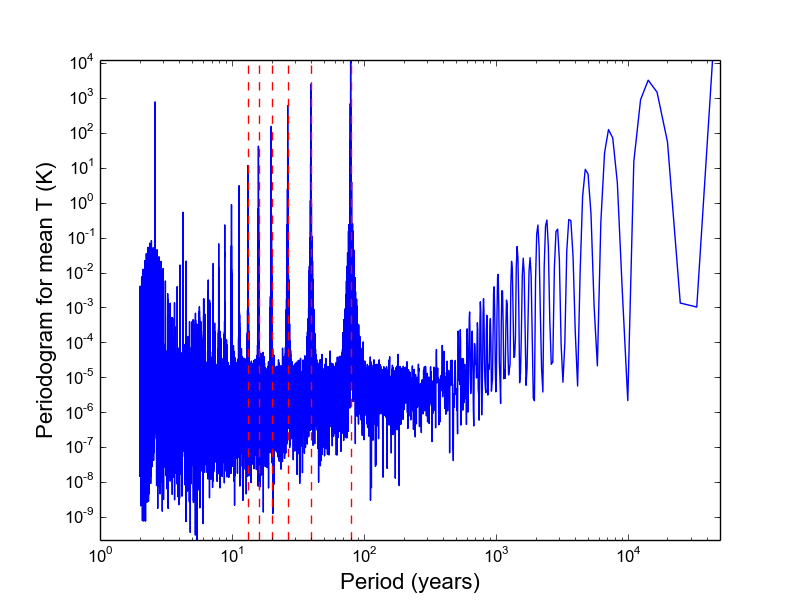} \\
\includegraphics[scale = 0.4]{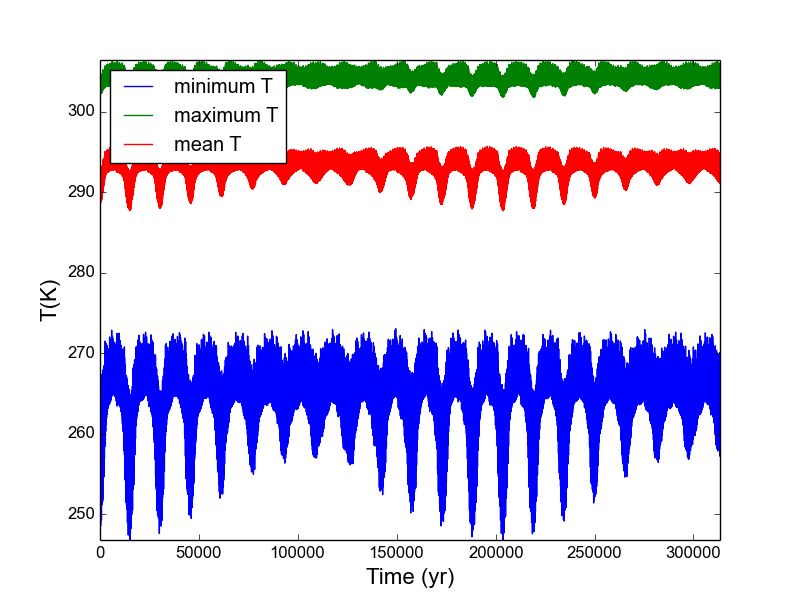} &
\includegraphics[scale=0.4]{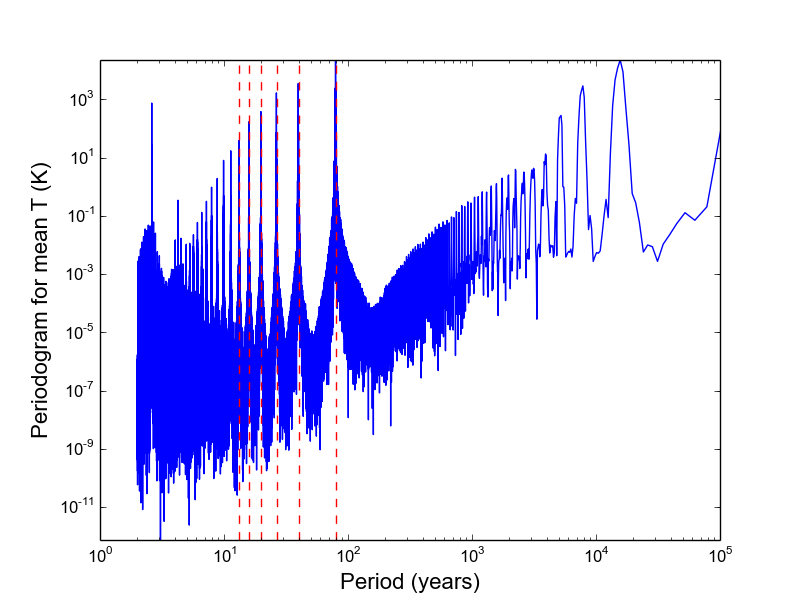} \\
\end{array}$
\caption{The climate evolution of the \alphacen B terrestrial planet, with obliquity evolution switched off (top row) and switched on (bottom row).  Left: The global maximum, minimum and mean temperatures on the surface over 100,000 years (obliquity evolution off) and over approximately 300,000 years (obliquity evolution on).  Right:  Periodograms for the mean temperature.  The red dashed lines indicate the binary's orbital period of 79 years, and its harmonics ($1/2$, $1/3$... of the period). \label{fig:alphacen_e0_climate}}
\end{center}
\end{figure*}

This phase drift results in markedly different climate evolution of the body, compared to the case where obliquity is held fixed (Figure \ref{fig:alphacen_e0_climate}).  In the fixed obliquity case, the eccentricity cycle induces a temperature oscillation of approximately 2K (to add to the radiative oscillation of 5K due to the changing proximity of \alphacen A).  The periodogram shows the two dominant oscillation modes at 79.9 and 14,500 years.  Their strength is indicated by the strength of their subsequent harmonics, which can be seen down to the tenth level!

A quite different picture emerges if obliquity evolution is activated (bottom row of Figure \ref{fig:alphacen_e0_climate}).  The temperature oscillations are now modulated by the phase drift between eccentricity and obliquity, which is periodic over $\sim$ 200,000 year timescales.  When the two cycles are in phase, we see the largest temperature oscillations (e.g. at $t\sim$ 200,000 years).  

\subsubsection{Adding a planet in 3:2 mean motion resonance}

\noindent We now consider joint planetary-binary Milankovitch cycles by adding an Earth mass planet on a circular orbit at 0.9293 AU, placing it in 3:2 mean motion resonance with the habitable planet.  Test runs with \alphacen A absent show the additional planet induces regular eccentricity oscillations in the habitable planet with amplitude of approximately 0.01, and a period of approximately 500 years.  Incidentally, the absence of \alphacen A would also place both planets outside the habitable zone.

\begin{figure*}
\begin{center}$\begin{array}{cc}
\includegraphics[scale=0.4]{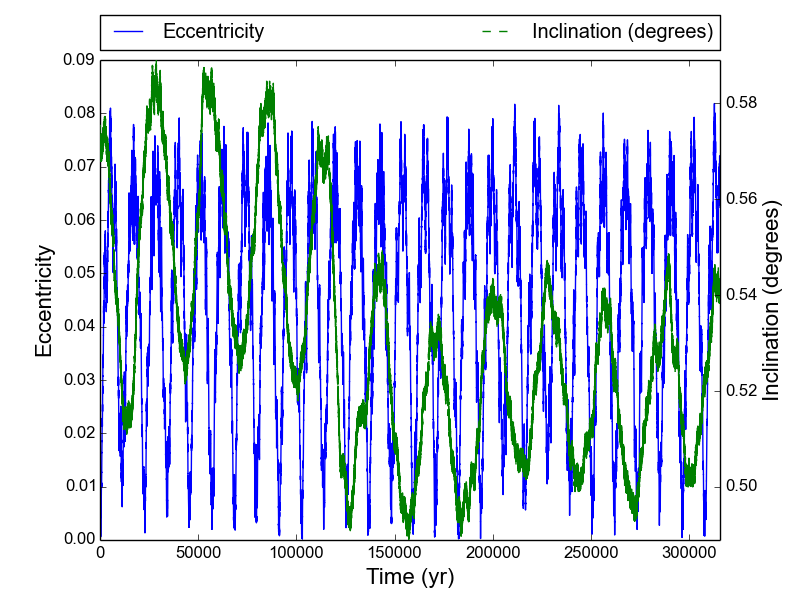} &
\includegraphics[scale=0.4]{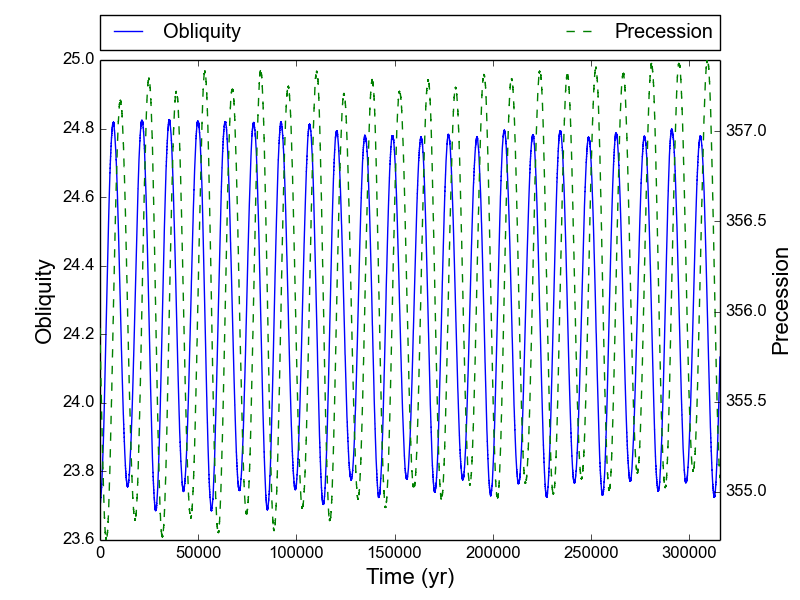}\\
\end{array}$
\caption{The dynamical evolution of the terrestrial planet orbiting \alphacen B. Left, the orbital evolution of the body, as given by its eccentricity and inclination.  Right, the spin evolution as given by the obliquity and precession angles. \label{fig:alphacen_e0_32_orbit}}
\end{center}
\end{figure*}

With \alphacen A present, the combination of stellar and planetary forcings produces eccentricity oscillations of maximum amplitude 0.08 (left panel of Figure \ref{fig:alphacen_e0_32_orbit}) and with a mix of dominant periods, as opposed to the distinct 14,500 year period observed in the single planet case.  The inclination varies with a period of approximately 30,000 years, with a distinctive shift in mean inclination of around 0.001 radians (i.e. 0.05$^\circ$).  The obliquity and precession continue to evolve at close to the eccentricity oscillation period, but the amplitude of their oscillations varies on approximately twice this timescale.

\begin{figure*}
\begin{center}$\begin{array}{cc}
\includegraphics[scale=0.4]{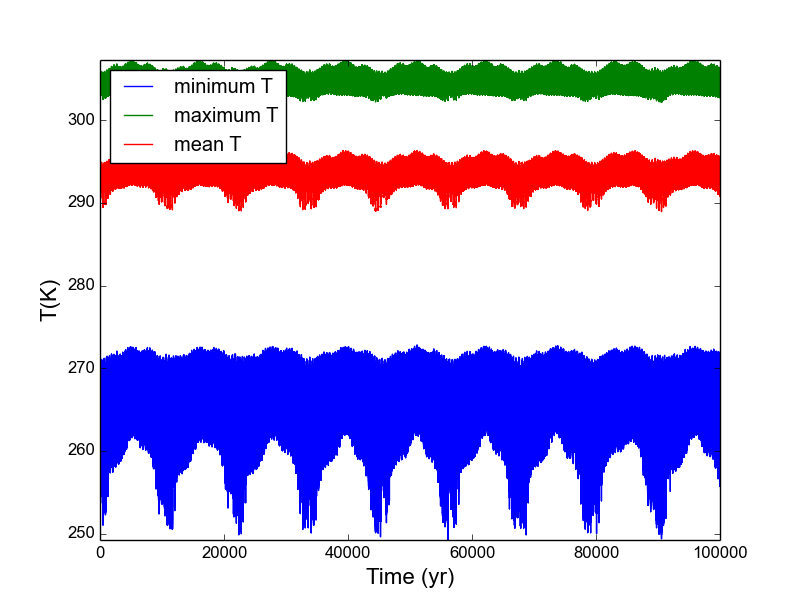} &
\includegraphics[scale=0.4]{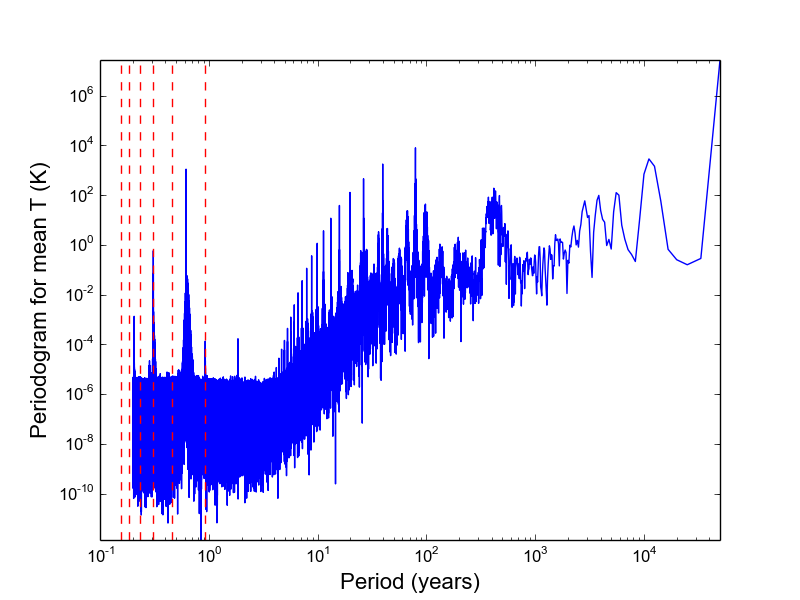} \\
\includegraphics[scale=0.4]{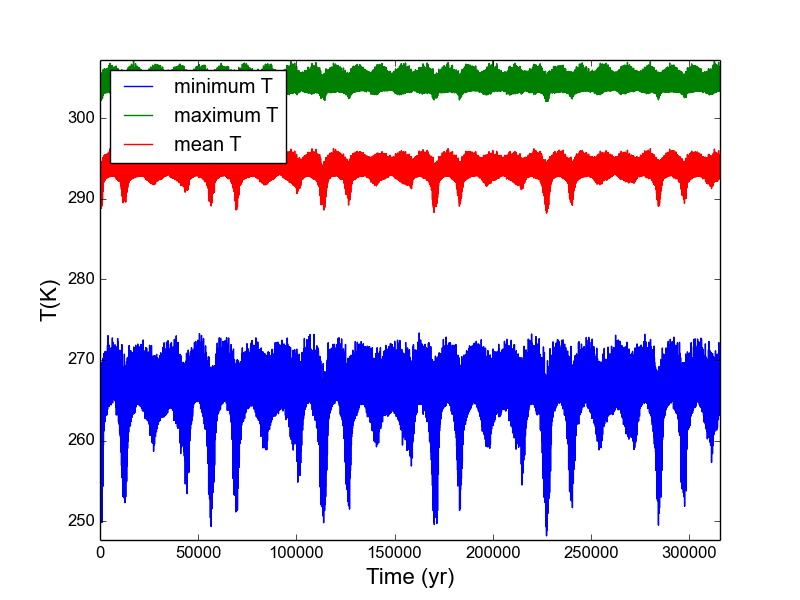} &
\includegraphics[scale=0.4]{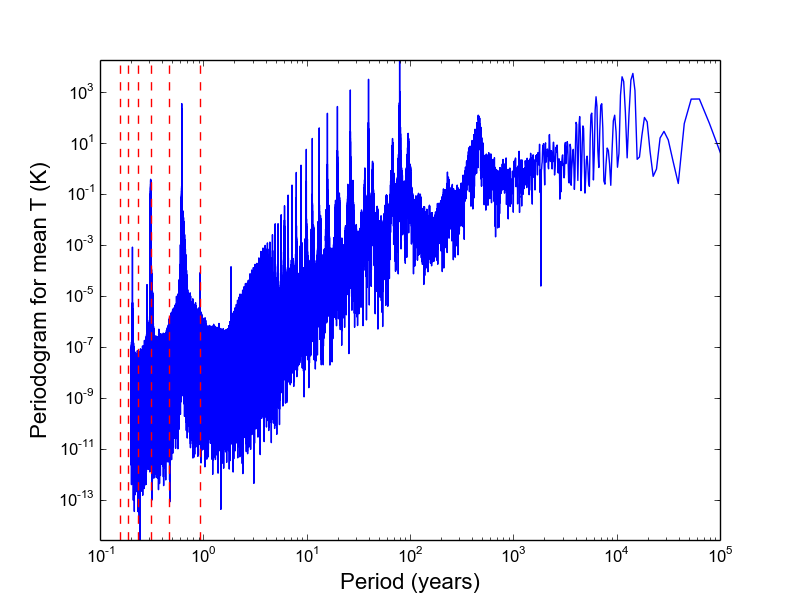} \\
\end{array}$
\caption{The climate evolution of the \alphacen B terrestrial planet, with obliquity evolution switched off (top row) and switched on (bottom row).  Left: The global maximum, minimum and mean temperatures on the surface over 100,000 years (obliquity evolution off) and over approximately 300,000 years (obliquity evolution on).  Right:  Periodograms for the mean temperature.  The red dashed lines indicate the binary's orbital period of 79 years, and its harmonics ($1/2$, $1/3$... of the period). \label{fig:alphacen_e0_32_climate}}
\end{center}
\end{figure*}

The uniform temperature evolution cycles seen in Figure \ref{fig:alphacen_e0_climate} are now more confused with the addition of a neighbour planet (Figure \ref{fig:alphacen_e0_32_climate}).  With obliquity evolution switched off (top row), the extra structure introduced into the eccentricity and inclination oscillations leaves an imprint on the temperature curves.  This can be seen in its periodogram (top right panel of Figure \ref{fig:alphacen_e0_32_climate}), which shows a relatively weak feature at the perturbing planet's orbital period, and at the resonant period of twice the perturber's period (or equivalently, three times the habitable planet's period).  The perturbations induced by the additional planet produce temperature variations of up to 2K compared to the single planet case.

With obliquity evolution turned on (bottom panel), the eccentricity/obliquity relationship seen in the previous case is preserved, resulting in phase drift between the two oscillations.  However, the extra structure in the eccentricity oscillation prevents the smooth amplitude modulation of temperature that we saw in the bottom right panel of Figure \ref{fig:alphacen_e0_climate}. It is broadly present, but heavily modified by the presence of the neighbouring planet. The periodogram still reveals weak signals at the perturbing planet's period, and the strong peak feature at approximately 14,500 years is now split in two.  There is also a significant increase in signal for periods of order 100-1000 years.    

Additional giant planets in a system like this might be expected to produce even larger excursions from circular orbits and stronger Milankovitch cycling.  Given that planet formation models disfavour the creation of Jupiter mass bodies in this system \citep{Xie2010} and are ruled out by observations of the \alphacen  system, at least at periods less than $\sim$ 1 year  \citep{Endl2001,Dumusque2012, Demory2015} this is not a particular concern.  But, one might imagine that undetected Neptune mass bodies could be present in this system on relatively long period orbits, and such bodies would be responsible for longer period Milankovitch cycles similar to that of Earth's.

\section{Discussion} \label{sec:Discussion}

\subsection{Limitations of the Model}

\noindent LEBM modelling is by its definition a compromise between the granularity of a climate simulation and computational expediency.  This compromise is stretched further by the coupling of the N-Body integrator and obliquity evolution.  We have adopted a very simple coupling where both the N-Body and LEBM components are constrained to follow the same global timestep.  

This timestep system works extremely well for systems where the dynamical timescale is relatively long, such as the S type binary systems.  In this scenario, the system timestep is limited only by the LEBM, and as such we can run simulations with similar wallclock times as that of a LEBM using fixed Keplerian orbits.  However, in the P type scenario, the dynamical timescale is relatively short, and the system is limited by the N Body timestep required to resolve the binary.  

There are several possible strategies for mitigating this timestep issue.  The most straightforward solution is to adopt a non-shared timestep for the N-Body component, allowing some of the bodies to possess shorter N Body timesteps.  This would reduce the computational load of evolving all the bodies (and the LEBM) at what can be very short timesteps.  Another solution would require the interpolation of body motions (in the case where the LEBM timestep is small compared to the N Body timestep), but this would likely produce only marginal gains in speed.  Perhaps the best solution for P type systems would be chain regularisation of the tight binary orbit \citep{Mikkola1990,Mikkola1993}.

Aside from the new challenges arising from the adoption of the N-Body integrator, there are the usual limitations that many LEBMs are subject to.  Our implementation of the LEBM is among the most simple available which can still broadly reproduce the seasonal temperature profiles of a fiducial Earth model.  The principal advantage of this simplicity is its ease of interpretation, but we must acknowledge that more advanced models may produce features we cannot.

For example, we do not model the carbonate-silicate (CS) cycle, which moderates fluctuations in atmospheric temperature by increasing and reducing the partial pressure of carbon dioxide.  The timescale on which we expect $CO_2$ levels to vary depends on the planet's geochemical properties, especially its ocean circulation.  For Earthlike planets, the equilbriation timescale of $CO_2$ is approximately half a million years \citep{Williams1997a} which is far shorter than the Milankovitch cycles experienced by the planetary bodies in this analysis.  However, our understanding of the CS cycle is rooted firmly in our understanding of the Earth, which orbits a single star.  It remains unclear whether a planet in a binary star system would possess a similar equilibriation timescale, even if the planet was effectively identical to the Earth.

While we have taken the first steps towards coupling celestial dynamics and LEBM climate modelling here, there are still several steps ahead of us.  For example, the tidal interactions between bodies will also modify orbits of habitable worlds, in particular reducing their eccentricity and modifying their rotational period \citep{Bolmont2014,Cunha2014}.  While this is unlikely to be an issue for the orbital configurations adopted in this analysis, it remains the case that while the tidal interactions between the binary stars is well characterised (e.g. \citealt{Mason2013, Zuluaga2016}), the tidal evolution of \emph{planets} in P type systems remains relatively unexplored.  

Also to be explored in full are the obliquity variations felt by planets in binary systems.  We have adopted a set of equations designed for a single star planetary system, and assumed they are valid when there are two stars present.  In effect we have assumed that in S type systems, the secondary's direct tidal torque on planetary spin is negligible, and that in P type systems the direct torques always co-add. Is this always the case? More investigation is needed.

We should also note that the strength of Milankovitch cycles measured by the LEBM will be an underestimate.  Tests conducted using Solar system parameters (Forgan \& Mead, in prep.) give Milankovitch cycles for the Earth that are an order of magnitude smaller in temperature variation than observed in paleoclimate data \citep{Zachos2001,Lisiecki2005}.  Paradoxically, stochastic EBMs, with additional random noise, can enhance periodic variations through the phenomenon of stochastic resonance \citep{Imkeller2001, Benzi2010}.  Obliquity variation does produce a much richer set of temperature variations on decadal timescales, which may be forced into stochastic resonant behaviour under appropriate circumstances.  Future investigations should consider adding a random noise term to the LEBM equation to permit this behaviour.
 
\subsection{Implications for Habitability}

So what have we gained by this coupling of N Body and LEBM integrators? Initially, we are able to confirm that in general, the decoupled approach of considering the radiative and gravitational perturbations separately is \emph{broadly} acceptable.  

Previous work in this field is not invalidated by our results, but it makes explicit some general principles that are already known implicitly.  Firstly, the habitable zone of a planetary system is defined by more than where the radiation sources are in the system.  The gravitational sources are equally important.  We know this on Earth thanks to our understanding of Milankovitch cycles, and the Earth's orbital and spin cycles are relatively weak when compared to measured cycles for Earthlike planets in typical exoplanet system configurations around a single star  (\citealt{Spiegel2010}, Forgan \& Mead, in prep.).

Secondly, the habitable zone of binary systems is even more sensitive to the gravitational field than single star systems.  This is already demonstrated implicitly by the N-Body simulations of orbital stability discussed in the Introduction.  Our results clearly identify the effect of orbital and spin stability on climate.  We show that relatively strong Milankovitch cycles exist in binary systems, even if there is only one planet present.  The periods of these cycles are in general shorter than that of single star systems, but of similar amplitudes.  Even on short timescales, the radiative perturbations induced over the orbital period of the binary are detectable in the mean temperature of the planet.

Thirdly, the circadian rhythms of life on planets in binary systems will be forced to adapt to the rhythms present in the binary system, as is evidenced by analogous studies of lunar photoperiodism in terrestrial organisms (\citealt{O'Malley2012a,Forgan2014b} and references within).  Temperature fluctuations of several K on timescales ranging from less than a year to almost a century (depending on whether the system is P or S type) is likely to produce significant fluctuations in surface coverage of biomes.  The rapid Milankovitch cycles are likely to play a stronger role also.  More sophisticated climate models coupled to N-Body physics (for example, 3D global circulation models) may show potential for more, shorter Ice Ages, and briefer interglacial periods.  The presence of such rapid changes to environmental selection pressure will have an indelible effect on the evolution of organisms in binary planetary systems.  Future work should build on recent attempts to produce 3D General Circulation Models of circumbinary planets (cf \citealt{May2016}),  incorporating the systems' gravitational evolution to determine these effects in detail.

\section{Conclusions }\label{sec:Conclusions}

\noindent We have investigated Milankovitch cycles both circumbinary (P type) and distant binary (S type) systems, using  Kepler-47 and $\alpha$ Centauri as archetypes.  To do this, we coupled a 1D latitudinal energy balance climate model (LEBM) with an N-Body integrator to follow the orbital evolution, and an obliquity evolution algorithm to study the spin-axis evolution.

We find that the combined spin-orbit-radiative perturbations induced by a companion star on a habitable planet produce Milankovitch cycles for both types of binary system, even when other planets are not present.  Periodogram analysis identifies both dynamical and secular oscillations in the mean temperature of planets in these systems, over a variety of short and long periods, as well as the presence of radiative perturbations directly linked to the period of the binary.  The strength of these oscillations is sensitive to the orbital configuration of the system.  The relative phase between eccentricity, precession and obliquity cycles is important, just as it is for the Earth.  

In general, we find these Milankovitch cycles are significantly shorter than comparable cycles on the Earth (in some cases shorter than 1000 years), although the amplitude of the changes they produce in the planets' orbital elements are comparable to those experienced by Earth.  This work demonstrates the need to consider joint dynamics-climate simulations of habitable worlds in binary systems, if we are to truly assess the potential for the birth and growth of biospheres on worlds with two suns.

\section*{Acknowledgments}

DHF gratefully acknowledges support from the ECOGAL project, grant agreement 291227, funded by the European Research Council under ERC-2011-ADG.  This work relied on the compute resources of the St Andrews MHD Cluster.  The author thanks both Nader Haghighipour and James Gilmore for insightful comments on an early version of this manuscript.   This research has  made use of NASA's Astrophysics Data System Bibliographic Services.  The code used in this work is now available open source as OBERON, which can be downloaded at \texttt{github.com/dh4gan/oberon}.

\bibliographystyle{mnras} 
\bibliography{nbody_EBM_binary}

\begin{thebibliography}{}
\makeatletter
\relax
\def\mn@urlcharsother{\let\do\@makeother \do\$\do\&\do\#\do\^\do\_\do\%\do\~}
\def\mn@doi{\begingroup\mn@urlcharsother \@ifnextchar [ {\mn@doi@}
  {\mn@doi@[]}}
\def\mn@doi@[#1]#2{\def\@tempa{#1}\ifx\@tempa\@empty \href
  {http://dx.doi.org/#2} {doi:#2}\else \href {http://dx.doi.org/#2} {#1}\fi
  \endgroup}
\def\mn@eprint#1#2{\mn@eprint@#1:#2::\@nil}
\def\mn@eprint@arXiv#1{\href {http://arxiv.org/abs/#1} {{\tt arXiv:#1}}}
\def\mn@eprint@dblp#1{\href {http://dblp.uni-trier.de/rec/bibtex/#1.xml}
  {dblp:#1}}
\def\mn@eprint@#1:#2:#3:#4\@nil{\def\@tempa {#1}\def\@tempb {#2}\def\@tempc
  {#3}\ifx \@tempc \@empty \let \@tempc \@tempb \let \@tempb \@tempa \fi \ifx
  \@tempb \@empty \def\@tempb {arXiv}\fi \@ifundefined
  {mn@eprint@\@tempb}{\@tempb:\@tempc}{\expandafter \expandafter \csname
  mn@eprint@\@tempb\endcsname \expandafter{\@tempc}}}

\bibitem[\protect\citeauthoryear{Anderson, Storch  \& Lai}{Anderson
  et~al.}{2016}]{Anderson2016}
Anderson K.~R.,  Storch N.~I.,   Lai D.,  2016, \mn@doi [MNRAS]
  {10.1093/mnras/stv2906}, 456, 3671

\bibitem[\protect\citeauthoryear{Andrade-Ines \& Michtchenko}{Andrade-Ines \&
  Michtchenko}{2014}]{Andrade-Ines2014}
Andrade-Ines E.,  Michtchenko T.~A.,  2014, \mn@doi [MNRAS]
  {10.1093/mnras/stu1591}, 444, 2167

\bibitem[\protect\citeauthoryear{Armstrong, Leovy  \& Quinn}{Armstrong
  et~al.}{2004}]{Armstrong2004}
Armstrong J.~C.,  Leovy C.~B.,   Quinn T.,  2004, \mn@doi [Icarus]
  {10.1016/j.icarus.2004.05.007}, 171, 255

\bibitem[\protect\citeauthoryear{Armstrong, Barnes, Domagal-Goldman, Breiner,
  Quinn  \& Meadows}{Armstrong et~al.}{2014a}]{Armstrong2014a}
Armstrong J.~C.,  Barnes R.,  Domagal-Goldman S.,  Breiner J.,  Quinn T.~R.,
  Meadows V.~S.,  2014a, \mn@doi [Astrobiology] {10.1089/ast.2013.1129}, 14,
  277

\bibitem[\protect\citeauthoryear{Armstrong, Osborn, Brown, Faedi, {Gomez Maqueo
  Chew}, Martin, Pollacco  \& Udry}{Armstrong et~al.}{2014b}]{Armstrong2014}
Armstrong D.~J.,  Osborn H.~P.,  Brown D. J.~A.,  Faedi F.,  {Gomez Maqueo
  Chew} Y.,  Martin D.~V.,  Pollacco D.,   Udry S.,  2014b, \mn@doi [MNRAS]
  {10.1093/mnras/stu1570}, 444, 1873

\bibitem[\protect\citeauthoryear{Benzi}{Benzi}{2010}]{Benzi2010}
Benzi R.,  2010, \mn@doi [Nonlinear Processes in Geophysics]
  {10.5194/npg-17-431-2010}, 17, 431

\bibitem[\protect\citeauthoryear{Bolmont, Raymond, von Paris, Selsis, Hersant,
  Quintana  \& Barclay}{Bolmont et~al.}{2014}]{Bolmont2014}
Bolmont E.,  Raymond S.~N.,  von Paris P.,  Selsis F.,  Hersant F.,  Quintana
  E.~V.,   Barclay T.,  2014, \mn@doi [ApJ] {10.1088/0004-637X/793/1/3}, 793, 3

\bibitem[\protect\citeauthoryear{Brown, Mead, Forgan, Raven  \& Cockell}{Brown
  et~al.}{2014}]{Brown2014}
Brown S.,  Mead A.,  Forgan D.,  Raven J.,   Cockell C.,  2014, \mn@doi
  [International Journal of Astrobiology] {10.1017/S1473550414000068}, 13, 279

\bibitem[\protect\citeauthoryear{Chavez, Georgakarakos, Prodan, Reyes-Ruiz,
  Aceves, Betancourt  \& Perez-Tijerina}{Chavez et~al.}{2014}]{Chavez2014}
Chavez C.~E.,  Georgakarakos N.,  Prodan S.,  Reyes-Ruiz M.,  Aceves H.,
  Betancourt F.,   Perez-Tijerina E.,  2014, \mn@doi [MNRAS]
  {10.1093/mnras/stu2142}, 446, 1283

\bibitem[\protect\citeauthoryear{Cuartas-Restrepo, Melita, Zuluaga, Portilla,
  Sucerquia  \& Miloni}{Cuartas-Restrepo et~al.}{2016}]{Cuartas-Restrepo2016}
Cuartas-Restrepo P.,  Melita M.,  Zuluaga J.,  Portilla B.,  Sucerquia M.,
  Miloni O.,  2016, MNRAS, pp submitted, eprint arXiv:1606.07546

\bibitem[\protect\citeauthoryear{Cunha, Correia  \& Laskar}{Cunha
  et~al.}{2014}]{Cunha2014}
Cunha D.,  Correia A.~C.,   Laskar J.,  2014, \mn@doi [International Journal of
  Astrobiology] {10.1017/S1473550414000226}, 14, 233

\bibitem[\protect\citeauthoryear{Cuntz}{Cuntz}{2014}]{Cuntz2014}
Cuntz M.,  2014, \mn@doi [ApJ] {10.1088/0004-637X/780/1/14}, 780, 14

\bibitem[\protect\citeauthoryear{{Del Genio}}{{Del Genio}}{1993}]{DelGenio1993}
{Del Genio} A.,  1993, \mn@doi [Icarus] {10.1006/icar.1993.1001}, 101, 1

\bibitem[\protect\citeauthoryear{{Del Genio}}{{Del Genio}}{1996}]{DelGenio1996}
{Del Genio} A.,  1996, \mn@doi [Icarus] {10.1006/icar.1996.0054}, 120, 332

\bibitem[\protect\citeauthoryear{Demory et~al.,}{Demory
  et~al.}{2015}]{Demory2015}
Demory B.-O.,  et~al., 2015, MNRAS, p. in press

\bibitem[\protect\citeauthoryear{Desidera \& Barbieri}{Desidera \&
  Barbieri}{2007}]{Desidera2007}
Desidera S.,  Barbieri M.,  2007, \mn@doi [A{\&}A]
  {10.1051/0004-6361:20066319}, 462, 345

\bibitem[\protect\citeauthoryear{Doolin \& Blundell}{Doolin \&
  Blundell}{2011}]{Doolin2011}
Doolin S.,  Blundell K.~M.,  2011, \mn@doi [MNRAS]
  {10.1111/j.1365-2966.2011.19657.x}, 418, 2656

\bibitem[\protect\citeauthoryear{Doyle et~al.,}{Doyle et~al.}{2011}]{Doyle2011}
Doyle L.~R.,  et~al., 2011, \mn@doi [Science (New York, N.Y.)]
  {10.1126/science.1210923}, 333, 1602

\bibitem[\protect\citeauthoryear{Dressing, Spiegel, Scharf, Menou  \&
  Raymond}{Dressing et~al.}{2010}]{Dressing2010}
Dressing C.~D.,  Spiegel D.~S.,  Scharf C.~A.,  Menou K.,   Raymond S.~N.,
  2010, \mn@doi [ApJ] {10.1088/0004-637X/721/2/1295}, 721, 1295

\bibitem[\protect\citeauthoryear{Dumusque et~al.,}{Dumusque
  et~al.}{2012}]{Dumusque2012}
Dumusque X.,  et~al., 2012, \mn@doi [Nature] {10.1038/nature11572}, 491, 207

\bibitem[\protect\citeauthoryear{Dunhill \& Alexander}{Dunhill \&
  Alexander}{2013}]{Dunhill2013}
Dunhill A.~C.,  Alexander R.~D.,  2013, \mn@doi [MNRAS]
  {10.1093/mnras/stt1456}, 435, 2328

\bibitem[\protect\citeauthoryear{Duquennoy \& Mayor}{Duquennoy \&
  Mayor}{1991}]{Duquennoy1991}
Duquennoy A.,  Mayor M.,  1991, A{\&}A, 248, 485

\bibitem[\protect\citeauthoryear{Dvorak}{Dvorak}{1984}]{Dvorak1984}
Dvorak R.,  1984, \mn@doi [Celestial Mechanics] {10.1007/BF01235815}, 34, 369

\bibitem[\protect\citeauthoryear{Dvorak}{Dvorak}{1986}]{Dvorak1986}
Dvorak R.,  1986, A{\&}A, 167, 379

\bibitem[\protect\citeauthoryear{Eggl, Pilat-Lohinger, Georgakarakos,
  Gyergyovits  \& Funk}{Eggl et~al.}{2012}]{Eggl2012}
Eggl S.,  Pilat-Lohinger E.,  Georgakarakos N.,  Gyergyovits M.,   Funk B.,
  2012, \mn@doi [ApJ] {10.1088/0004-637X/752/1/74}, 752, 74

\bibitem[\protect\citeauthoryear{Endl, K�rster, Els, Hatzes  \& Cochran}{Endl
  et~al.}{2001}]{Endl2001}
Endl M.,  K�rster M.,  Els S.,  Hatzes A.~P.,   Cochran W.~D.,  2001, \mn@doi
  [Astronomy and Astrophysics] {10.1051/0004-6361:20010723}, 374, 675

\bibitem[\protect\citeauthoryear{Farrell}{Farrell}{1990}]{Farrell1990}
Farrell B.~F.,  1990, \mn@doi [Journal of Atmospheric Sciences]
  {10.1175/1520-0469(1990)047&lt;2986:ECD&gt;2.0.CO;2}, 47, 2986

\bibitem[\protect\citeauthoryear{Forgan}{Forgan}{2012}]{Forgan2012}
Forgan D.,  2012, \mn@doi [MNRAS] {10.1111/j.1365-2966.2012.20698.x}, 422, 1241

\bibitem[\protect\citeauthoryear{Forgan}{Forgan}{2014}]{Forgan2014}
Forgan D.,  2014, \mn@doi [MNRAS] {10.1093/mnras/stt1964}, 437, 1352

\bibitem[\protect\citeauthoryear{Forgan, Mead, Cockell  \& Raven}{Forgan
  et~al.}{2015}]{Forgan2014b}
Forgan D.~H.,  Mead A.,  Cockell C.~S.,   Raven J.~A.,  2015, \mn@doi
  [International Journal of Astrobiology] {10.1017/S147355041400041X}, 14, 465

\bibitem[\protect\citeauthoryear{{G. Martin}, Armitage  \& Alexander}{{G.
  Martin} et~al.}{2013}]{Martin2013}
{G. Martin} R.,  Armitage P.~J.,   Alexander R.~D.,  2013, \mn@doi [ApJ]
  {10.1088/0004-637X/773/1/74}, 773, 74

\bibitem[\protect\citeauthoryear{Haghighipour \& Kaltenegger}{Haghighipour \&
  Kaltenegger}{2013}]{Haghighipour2013}
Haghighipour N.,  Kaltenegger L.,  2013, \mn@doi [ApJ]
  {10.1088/0004-637X/777/2/166}, 777, 166

\bibitem[\protect\citeauthoryear{Hart}{Hart}{1979}]{Hart_HZ}
Hart M.,  1979, \mn@doi [Icarus] {10.1016/0019-1035(79)90141-6}, 37, 351

\bibitem[\protect\citeauthoryear{Hatzes}{Hatzes}{2013}]{Hatzes2013}
Hatzes A.~P.,  2013, \mn@doi [ApJ] {10.1088/0004-637X/770/2/133}, 770, 133

\bibitem[\protect\citeauthoryear{Hatzes, Cochran, Endl, McArthur, Paulson,
  Walker, Campbell  \& Yang}{Hatzes et~al.}{2003}]{Hatzes2003}
Hatzes A.~P.,  Cochran W.~D.,  Endl M.,  McArthur B.,  Paulson D.~B.,  Walker
  G. A.~H.,  Campbell B.,   Yang S.,  2003, \mn@doi [ApJ] {10.1086/379281},
  599, 1383

\bibitem[\protect\citeauthoryear{Holman \& Wiegert}{Holman \&
  Wiegert}{1999}]{Holman1999a}
Holman M.~J.,  Wiegert P.~A.,  1999, \mn@doi [The Astronomical Journal]
  {10.1086/300695}, 117, 621

\bibitem[\protect\citeauthoryear{Huang}{Huang}{1959}]{Huang1959}
Huang S.-S.,  1959, \mn@doi [PASP] {10.1086/127417}, 71, 421

\bibitem[\protect\citeauthoryear{Imkeller}{Imkeller}{2001}]{Imkeller2001}
Imkeller P.,  2001, in Imkeller P.,  {Von Storch} J.-S.,  eds, , Stochastic
  Climate Models.
Birkh{\"{a}}user Basel, Basel, pp 213--240,
  \mn@doi{10.1007/978-3-0348-8287-3_9}, \url
  {http://link.springer.com/10.1007/978-3-0348-8287-3{\_}9}

\bibitem[\protect\citeauthoryear{Jaime, Aguilar  \& Pichardo}{Jaime
  et~al.}{2014}]{Jaime2014}
Jaime L.~G.,  Aguilar L.,   Pichardo B.,  2014, \mn@doi [MNRAS]
  {10.1093/mnras/stu1052}, 443, 260

\bibitem[\protect\citeauthoryear{Kaltenegger \& Haghighipour}{Kaltenegger \&
  Haghighipour}{2013}]{Kaltenegger2013}
Kaltenegger L.,  Haghighipour N.,  2013, \mn@doi [ApJ]
  {10.1088/0004-637X/777/2/165}, 777, 165

\bibitem[\protect\citeauthoryear{Kane \& Hinkel}{Kane \&
  Hinkel}{2013}]{Kane2013}
Kane S.~R.,  Hinkel N.~R.,  2013, \mn@doi [ApJ] {10.1088/0004-637X/762/1/7},
  762, 7

\bibitem[\protect\citeauthoryear{Kasting, Whitmire  \& Reynolds}{Kasting
  et~al.}{1993}]{Kasting_et_al_93}
Kasting J.,  Whitmire D.,   Reynolds R.,  1993, \mn@doi [Icarus]
  {10.1006/icar.1993.1010}, 101, 108

\bibitem[\protect\citeauthoryear{Kopparapu et~al.,}{Kopparapu
  et~al.}{2013}]{Kopparapu2013}
Kopparapu R.~K.,  et~al., 2013, \mn@doi [ApJ] {10.1088/0004-637X/765/2/131},
  765, 131

\bibitem[\protect\citeauthoryear{Kopparapu, Ramirez, SchottelKotte, Kasting,
  Domagal-Goldman  \& Eymet}{Kopparapu et~al.}{2014}]{Kopparapu2014}
Kopparapu R.~K.,  Ramirez R.~M.,  SchottelKotte J.,  Kasting J.~F.,
  Domagal-Goldman S.,   Eymet V.,  2014, \mn@doi [ApJ]
  {10.1088/2041-8205/787/2/L29}, 787, L29

\bibitem[\protect\citeauthoryear{Kostov et~al.,}{Kostov
  et~al.}{2014}]{Kostov2014}
Kostov V.~B.,  et~al., 2014, \mn@doi [ApJ] {10.1088/0004-637X/784/1/14}, 784,
  14

\bibitem[\protect\citeauthoryear{Laskar}{Laskar}{1986a}]{Laskar1986}
Laskar J.,  1986a, A{\&}A, 157, 59

\bibitem[\protect\citeauthoryear{Laskar}{Laskar}{1986b}]{Laskar1986a}
Laskar J.,  1986b, A{\&}A, 164, 437

\bibitem[\protect\citeauthoryear{Lisiecki \& Raymo}{Lisiecki \&
  Raymo}{2005}]{Lisiecki2005}
Lisiecki L.~E.,  Raymo M.~E.,  2005, \mn@doi [Paleoceanography]
  {10.1029/2004PA001071}, 20, id PA1003

\bibitem[\protect\citeauthoryear{Marzari, Thebault, Scholl, Picogna  \&
  Baruteau}{Marzari et~al.}{2013}]{Marzari2013}
Marzari F.,  Thebault P.,  Scholl H.,  Picogna G.,   Baruteau C.,  2013,
  \mn@doi [Astronomy {\&} Astrophysics] {10.1051/0004-6361/201220893}, 553, A71

\bibitem[\protect\citeauthoryear{Mason, Zuluaga, Clark  \&
  Cuartas-Restrepo}{Mason et~al.}{2013}]{Mason2013}
Mason P.~a.,  Zuluaga J.~I.,  Clark J.~M.,   Cuartas-Restrepo P.~a.,  2013,
  \mn@doi [ApJ] {10.1088/2041-8205/774/2/L26}, 774, L26

\bibitem[\protect\citeauthoryear{May \& Rauscher}{May \&
  Rauscher}{2016}]{May2016}
May E.~M.,  Rauscher E.,  2016, \mn@doi [The Astrophysical Journal]
  {10.3847/0004-637X/826/2/225}, 826, 225

\bibitem[\protect\citeauthoryear{Meschiari}{Meschiari}{2014}]{Meschiari2014}
Meschiari S.,  2014, \mn@doi [ApJ] {10.1088/0004-637X/790/1/41}, 790, 41

\bibitem[\protect\citeauthoryear{Mikkola \& Aarseth}{Mikkola \&
  Aarseth}{1990}]{Mikkola1990}
Mikkola S.,  Aarseth S.~J.,  1990, Celestial Mechanics and Dynamical Astronomy
  (ISSN 0923-2958), 47, 375

\bibitem[\protect\citeauthoryear{Mikkola \& Aarseth}{Mikkola \&
  Aarseth}{1993}]{Mikkola1993}
Mikkola S.,  Aarseth S.~J.,  1993, \mn@doi [Celestial Mechanics {\&} Dynamical
  Astronomy] {10.1007/BF00695714}, 57, 439

\bibitem[\protect\citeauthoryear{O'Malley-James, Raven, Cockell  \&
  Greaves}{O'Malley-James et~al.}{2012}]{O'Malley2012a}
O'Malley-James J.~T.,  Raven J.~A.,  Cockell C.~S.,   Greaves J.~S.,  2012,
  \mn@doi [Astrobiology] {10.1089/ast.2011.0678}, 12, 115

\bibitem[\protect\citeauthoryear{Orosz et~al.,}{Orosz et~al.}{2012}]{Orosz2012}
Orosz J.~A.,  et~al., 2012, \mn@doi [Science (New York, N.Y.)]
  {10.1126/science.1228380}, 337, 1511

\bibitem[\protect\citeauthoryear{Pichardo, Sparke  \& Aguilar}{Pichardo
  et~al.}{2005}]{Pichardo2005}
Pichardo B.,  Sparke L.~S.,   Aguilar L.~A.,  2005, \mn@doi [MNRAS]
  {10.1111/j.1365-2966.2005.08905.x}, 359, 521

\bibitem[\protect\citeauthoryear{Pierrehumbert}{Pierrehumbert}{2005}]{Pierrehu%
mbert2005}
Pierrehumbert R.~T.,  2005, \mn@doi [Journal of Geophysical Research]
  {10.1029/2004JD005162}, 110, D01111

\bibitem[\protect\citeauthoryear{Pilat-Lohinger \& Dvorak}{Pilat-Lohinger \&
  Dvorak}{2002}]{Pilat-Lohinger2002}
Pilat-Lohinger E.,  Dvorak R.,  2002, \mn@doi [Celestial Mechanics and
  Dynamical Astronomy] {10.1023/A:1014586308539}, pp 143--153

\bibitem[\protect\citeauthoryear{Queloz et~al.,}{Queloz
  et~al.}{2000}]{Queloz2000}
Queloz D.,  et~al., 2000, A{\&}A, 354, 99

\bibitem[\protect\citeauthoryear{Quintana, Lissauer, Chambers  \&
  Duncan}{Quintana et~al.}{2002}]{Quintana2002}
Quintana E.~V.,  Lissauer J.~J.,  Chambers J.~E.,   Duncan M.~J.,  2002,
  \mn@doi [ApJ] {10.1086/341808}, 576, 982

\bibitem[\protect\citeauthoryear{Quintana, Adams, Lissauer  \&
  Chambers}{Quintana et~al.}{2007}]{Quintana2007}
Quintana E.~V.,  Adams F.~C.,  Lissauer J.~J.,   Chambers J.~E.,  2007, \mn@doi
  [ApJ] {10.1086/512542}, 660, 807

\bibitem[\protect\citeauthoryear{Rafikov}{Rafikov}{2013}]{Rafikov2013}
Rafikov R.~R.,  2013, \mn@doi [ApJ] {10.1088/2041-8205/764/1/L16}, 764, L16

\bibitem[\protect\citeauthoryear{Rafikov \& Silsbee}{Rafikov \&
  Silsbee}{2014a}]{Rafikov2014a}
Rafikov R.~R.,  Silsbee K.,  2014a, \mn@doi [ApJ] {10.1088/0004-637X/798/2/69},
  798, 69

\bibitem[\protect\citeauthoryear{Rafikov \& Silsbee}{Rafikov \&
  Silsbee}{2014b}]{Rafikov2014}
Rafikov R.~R.,  Silsbee K.,  2014b, \mn@doi [ApJ] {10.1088/0004-637X/798/2/70},
  798, 70

\bibitem[\protect\citeauthoryear{Raghavan et~al.,}{Raghavan
  et~al.}{2010}]{Raghavan2010}
Raghavan D.,  et~al., 2010, ApJSS, 190, 1

\bibitem[\protect\citeauthoryear{Rajpaul, Aigrain  \& Roberts}{Rajpaul
  et~al.}{2016}]{Rajpaul2015}
Rajpaul V.,  Aigrain S.,   Roberts S.,  2016, \mn@doi [MNRAS]
  {10.1093/mnrasl/slv164}, 456, L6

\bibitem[\protect\citeauthoryear{Sigurdsson, Richer, Hansen, Stairs  \&
  Thorsett}{Sigurdsson et~al.}{2003}]{Sigurdsson2003}
Sigurdsson S.,  Richer H.~B.,  Hansen B.~M.,  Stairs I.~H.,   Thorsett S.~E.,
  2003, \mn@doi [Science (New York, N.Y.)] {10.1126/science.1086326}, 301, 193

\bibitem[\protect\citeauthoryear{Silsbee \& Rafikov}{Silsbee \&
  Rafikov}{2015}]{Silsbee2015}
Silsbee K.,  Rafikov R.~R.,  2015, \mn@doi [ApJ] {10.1088/0004-637X/798/2/71},
  798, 71

\bibitem[\protect\citeauthoryear{Spiegel, Menou  \& Scharf}{Spiegel
  et~al.}{2008}]{Spiegel_et_al_08}
Spiegel D.~S.,  Menou K.,   Scharf C.~A.,  2008, \mn@doi [ApJ]
  {10.1086/588089}, 681, 1609

\bibitem[\protect\citeauthoryear{Spiegel, Raymond, Dressing, Scharf  \&
  Mitchell}{Spiegel et~al.}{2010}]{Spiegel2010}
Spiegel D.~S.,  Raymond S.~N.,  Dressing C.~D.,  Scharf C.~A.,   Mitchell
  J.~L.,  2010, \mn@doi [ApJ] {10.1088/0004-637X/721/2/1308}, 721, 1308

\bibitem[\protect\citeauthoryear{Tajika}{Tajika}{2008}]{Tajika2008}
Tajika E.,  2008, \mn@doi [ApJ] {10.1086/589831}, 680, L53

\bibitem[\protect\citeauthoryear{Th{\'{e}}bault, Marzari  \&
  Scholl}{Th{\'{e}}bault et~al.}{2008}]{Thebault2008}
Th{\'{e}}bault P.,  Marzari F.,   Scholl H.,  2008, \mn@doi [MNRAS]
  {10.1111/j.1365-2966.2008.13536.x}, 388, 1528

\bibitem[\protect\citeauthoryear{Th{\'{e}}bault, Marzari  \&
  Scholl}{Th{\'{e}}bault et~al.}{2009}]{Thebault2009}
Th{\'{e}}bault P.,  Marzari F.,   Scholl H.,  2009, \mn@doi [MNRAS]
  {10.1111/j.1745-3933.2008.00590.x}, 393, L21

\bibitem[\protect\citeauthoryear{Thevenin, Provost, Morel, Berthomieu, Bouchy
  \& Carrier}{Thevenin et~al.}{2002}]{Thevenin2002}
Thevenin F.,  Provost J.,  Morel P.,  Berthomieu G.,  Bouchy F.,   Carrier F.,
  2002, \mn@doi [Astronomy and Astrophysics] {10.1051/0004-6361:20021074}, 392,
  L9

\bibitem[\protect\citeauthoryear{Thorsett, Arzoumanian  \& Taylor}{Thorsett
  et~al.}{1993}]{Thorsett1993}
Thorsett S.~E.,  Arzoumanian Z.,   Taylor J.~H.,  1993, \mn@doi [ApJ]
  {10.1086/186933}, 412, L33

\bibitem[\protect\citeauthoryear{Vladilo, Murante, Silva, Provenzale, Ferri  \&
  Ragazzini}{Vladilo et~al.}{2013}]{Vladilo2013}
Vladilo G.,  Murante G.,  Silva L.,  Provenzale A.,  Ferri G.,   Ragazzini G.,
  2013, \mn@doi [ApJ] {10.1088/0004-637X/767/1/65}, 767, 65

\bibitem[\protect\citeauthoryear{Welsh et~al.,}{Welsh et~al.}{2012}]{Welsh2012}
Welsh W.~F.,  et~al., 2012, \mn@doi [Nature] {10.1038/nature10768}, 481, 475

\bibitem[\protect\citeauthoryear{Welsh et~al.,}{Welsh et~al.}{2015}]{Welsh2014}
Welsh W.~F.,  et~al., 2015, \mn@doi [ApJ] {10.1088/0004-637X/809/1/26}, 809, 26

\bibitem[\protect\citeauthoryear{Wertheimer \& Laughlin}{Wertheimer \&
  Laughlin}{2006}]{Wertheimer2006}
Wertheimer J.~G.,  Laughlin G.,  2006, \mn@doi [The Astronomical Journal]
  {10.1086/507771}, 132, 1995

\bibitem[\protect\citeauthoryear{Wiegert \& Holman}{Wiegert \&
  Holman}{1997}]{Wiegert1997}
Wiegert P.~A.,  Holman M.~J.,  1997, \mn@doi [The Astronomical Journal]
  {10.1086/118360}, 113, 1445

\bibitem[\protect\citeauthoryear{Williams \& Kasting}{Williams \&
  Kasting}{1997}]{Williams1997a}
Williams D.,  Kasting J.,  1997, \mn@doi [Icarus] {10.1006/icar.1997.5759},
  129, 254

\bibitem[\protect\citeauthoryear{Xie, Zhou  \& Ge}{Xie et~al.}{2010}]{Xie2010}
Xie J.-W.,  Zhou J.-L.,   Ge J.,  2010, \mn@doi [ApJ]
  {10.1088/0004-637X/708/2/1566}, 708, 1566

\bibitem[\protect\citeauthoryear{Zachos, Pagani, Sloan, Thomas  \&
  Billups}{Zachos et~al.}{2001}]{Zachos2001}
Zachos J.,  Pagani M.,  Sloan L.,  Thomas E.,   Billups K.,  2001, \mn@doi
  [Science (New York, N.Y.)] {10.1126/science.1059412}, 292, 686

\bibitem[\protect\citeauthoryear{Zucker, Mazeh, Santos, Udry  \& Mayor}{Zucker
  et~al.}{2004}]{Zucker2004}
Zucker S.,  Mazeh T.,  Santos N.~C.,  Udry S.,   Mayor M.,  2004, \mn@doi
  [A{\&}A] {10.1051/0004-6361:20040384}, 426, 695

\bibitem[\protect\citeauthoryear{Zuluaga, Mason  \& Cuartas-Restrepo}{Zuluaga
  et~al.}{2016}]{Zuluaga2016}
Zuluaga J.~I.,  Mason P.~A.,   Cuartas-Restrepo P.~A.,  2016, \mn@doi [The
  Astrophysical Journal] {10.3847/0004-637X/818/2/160}, 818, 160

\makeatother
\end{thebibliography}

\appendix

\label{lastpage}

\end{document}